# Artificial Intelligence for Direct Prediction of Molecular Dynamics Across Chemical Space

Fuchun Ge,[1] Yuxinxin Chen,[1,3] and Pavlo O. Dral[1,2,3*]

[1]*State Key Laboratory of Physical Chemistry of Solid Surfaces, Department of Chemistry, College of Chemistry and Chemical Engineering, and Fujian Provincial Key Laboratory of Theoretical and Computational Chemistry, Xiamen University, Xiamen 361005, China.*

[2]*Institute of Physics, Faculty of Physics, Astronomy, and Informatics, Nicolaus Copernicus University in Toruń, ul. Grudziądzka 5, 87-100 Toruń, Poland.*

[3]*Aitomistic, Shenzhen 518000, China.*

*Email:* dral@xmu.edu.cn

**Abstract**

Molecular dynamics (MD) is a powerful tool for exploring the behavior of atomistic systems, but its reliance on sequential numerical integration limits simulation efficiency. We present a novel neural network architecture, MDtrajNet, and a pre-trained foundational model, MDtrajNet-1, that directly generates MD trajectories across chemical space, bypassing force calculations and integration. MDtrajNet combines equivariant neural networks with a transformer-based architecture to achieve strong accuracy and transferability in predicting long-time trajectories. This approach accelerates simulations by up to two orders of magnitude and yields better accuracy than MD propagated with established machine-learning interatomic potentials trained on the same data. Remarkably, the errors of the trajectories generated by MDtrajNet-1 for various seen and even unseen small-sized molecular systems are close to those of the conventional *ab initio* MD. The current limitations of MDtrajNet-1 are attributed to the relatively small size of the chemical space in its training data; however, even for bigger, unseen systems, MDtrajNet-1 provides a good starting point for fine-tuning and obtaining system-specific models. The architecture's flexible design supports diverse application scenarios, including different statistical ensembles, boundary conditions, and interaction types. By overcoming the intrinsic speed barrier of conventional MD, MDtrajNet opens new frontiers in efficient and scalable atomistic simulations.

## Introduction

Molecules are inherently dynamic entities, constantly in motion due to quantum mechanical (QM) interactions.[1] Accurately simulating this motion is essential for understanding molecular properties and behaviors beyond what can be inferred from a single, static configuration obtained through conventional QM calculations. Molecular dynamics (MD) simulations serve this purpose by modeling the time evolution of molecular systems through trajectories—sequences of nuclear positions generated at discrete time steps.[2-4]

This trajectory-based framework has become indispensable in numerous applications, including molecular ensemble generation, spectroscopy simulation, conformational sampling, reaction mechanism elucidation, and protein folding studies.[5-9] However, the high computational cost of MD remains a significant limitation. The need for extremely small time steps and long simulation durations results in an enormous number of iterative steps, each requiring expensive force evaluations. This cost severely constrains the size of molecular systems and the timescales that can be practically simulated, especially when QM-level accuracy is desired, i.e., in *ab initio* MD (AIMD).[10, 11]

Machine learning interatomic potentials (MLIPs), which are trained to approximate QM-level forces, have emerged as a powerful strategy to accelerate MD simulations (Figure 1b). [10, 12-14]While MLIPs greatly reduce the per-step cost, the iterative nature of MD—fundamentally non-parallelizable—remains a core bottleneck, limiting overall efficiency.

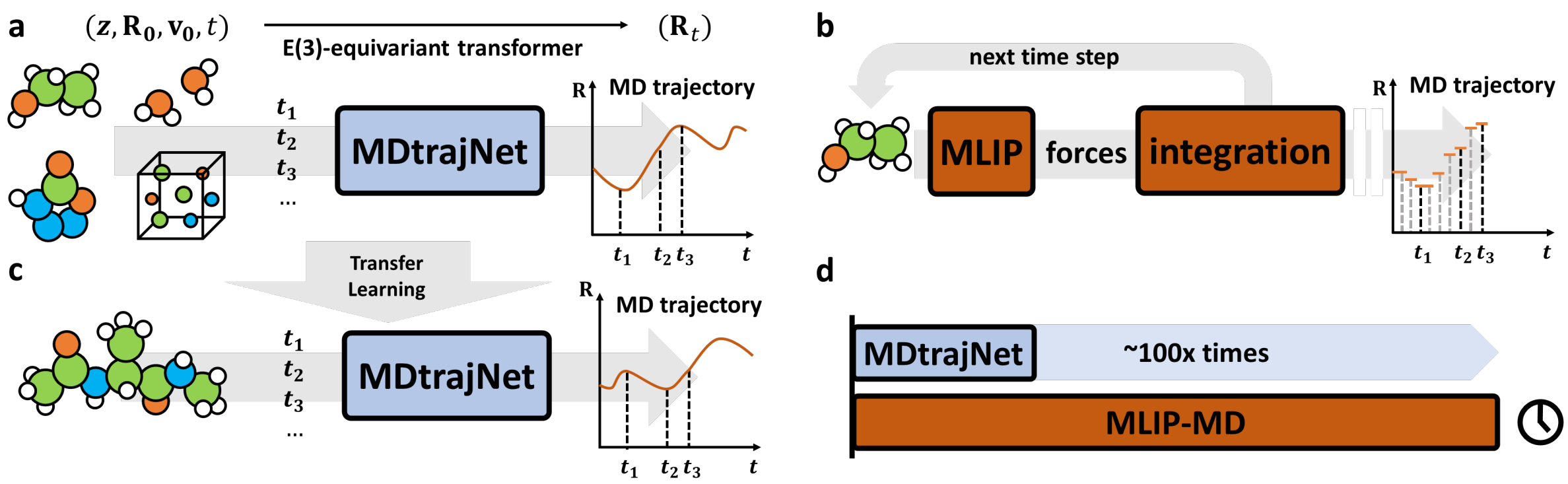

**Figure 1.** The novel neural network architecture MDtrajNet introduced here conceptually compared to machine learning interatomic potentials (MLIPs). a) MDtrajNet is based on an E(3)-equivariant network combined with the transformer architecture, imparting it the ability to directly predict nuclear positions at the required time; it works for different molecules and can be applied to materials simulations. b) In contrast, MLIPs only accelerate the force calculations, which do not solve the fundamental bottleneck of MD, which is an iterative, non-parallelizable trajectory propagation. c) After fine-tuning, MDtrajNet can be used to simulate increasingly complex systems. d) MDtrajNet achieves faster MD simulation speed than using MLIPs.

To address this intrinsic limitation, we propose a fundamentally different approach: directly predicting nuclear positions as a function of time, bypassing iterative force-based propagation entirely. To date, only one such model has been published: our proof-of-concept four-dimensional spacetime (4D) model, GICnet[15], which directly predicts short MD trajectories. The term "4D-spacetime" reflects the model's capacity to learn molecular motion in three spatial dimensions plus time. Given a molecule's initial conditions and a target time point, GICnet predicts future geometries and velocities without solving the equations of motion step-by-step. This novel approach enables accurate trajectory propagation without force calculations, achieving order-of-magnitude efficiency improvements over MLIP-accelerated MD while preserving long-term physical fidelity, as demonstrated across various molecules. We note that our 4D concept is distinct from emerging alternative approaches accelerating MD by propagating with longer strides:[16-18] 4D models are also complementary to the emerging deep generative models, which focus on stochastic sampling of the distribution of configurations.[19] 4D-spacetime models are smooth, deterministic functions of time, equally suitable for smooth interpolation between time steps and propagating with longer strides. 4D models also provide unique insight into chemical reactions as they become analytical expressions contained in the neural network.[15]

Despite its promise, GICnet has a critical limitation: it is not transferable beyond the specific molecule it was trained on. This lack of generality arises from the use of generalized internal coordinates as descriptors and a simple dense neural network (NN) architecture, which imposes severe constraints for real-world applications. A fundamentally new model architecture is therefore required to develop a transferable, general-purpose 4D-spacetime model.

To this end, we identified the transformer architecture[20] as a promising foundation. Transformers naturally handle variable-length inputs and, when used without positional encoding, preserve permutation invariance or equivariance—key properties for building transferable models for direct MD trajectory prediction. Transformer-based generative models have revolutionized fields such as natural language processing[21, 22] and computer vision[23-25], largely due to their attention mechanism, which enables dynamic correlation analysis across input components. In the molecular context, this capability maps well to atomic interactions. Notably, transformers have already been applied to tasks such as geometry optimization, conformer generation, and molecular representation learning in computational chemistry.[26-28]

Another promising direction is the use of equivariant neural networks (equivariant NNs), which preserve symmetry-group equivariance from input to output. Equivariance is a powerful inductive bias and has proven effective across domains—from image recognition to the growing class of MLIPs (e.g., MACE[29], NequIP[30], Allegro[31], EGNN[32], PaiNN[33], NewtonNet[34]). Although energy prediction in MLIPs requires only E(3)-invariance (the special case of equivariance for scalars), using equivariant features internally has been shown to significantly enhance model performance and data efficiency. For 4D-spacetime models, where output vectors represent positions, equivariant NNs are particularly well-suited to ensure physical consistency under E(3) transformations.

The integration of transformer architectures with equivariance is gaining traction across multiple fields, including 2D/3D pattern recognition[35, 36], molecular potential modeling[37, 38], and protein structure prediction[39]. For instance, SE(3)-transformers[40] have been used to predict the positions of charged particles under electrostatic interactions. While these models currently do not incorporate time as an explicit input—limiting them to single time-step predictions—they demonstrate strong potential for learning the dynamics of multi-body systems, including molecular systems.

In this work, we build upon our original 4D-spacetime concept and introduce MDtrajNet, a novel architecture combining transformers with equivariant NNs. MDtrajNet enables MD simulations that are several orders of magnitude faster than even MLIP-accelerated MD, while providing trajectories of higher accuracy (Figure 1d). We used this architecture to train a foundational model, MDtrajNet-1, on a newly introduced, diverse dataset. MDtrajNet-1 is designed for broad transferability across chemical compound space (Figure 1a). We show that the pre-trained weights of MDtrajNet-1 facilitate efficient fine-tuning, allowing rapid adaptation to increasingly complex molecular systems (Figure 1c).

## Results and discussion

### *Neural network architecture*

MDtrajNet takes the initial conditions (initial nuclear positions $\mathbf{R}_0$ and velocities $\mathbf{v}_0$) and a target time $t$ as inputs, and predicts the molecular structure coordinates $\mathbf{R}_t$ at time $t$, thereby establishing a connection between two frames on the trajectory. Importantly, MDtrajNet is designed to be transferable across systems containing diverse atomic species: as the result, MDtrajNet also requires the corresponding element-type information encoded with a vector of atomic numbers $\boldsymbol{z}$. Bringing it all together, the MDtrajNet is a function of the following form:

$$\mathbf{R}_t = f_{\text{MDtrajNet}}(\mathbf{z}, \mathbf{R}_0, \mathbf{v}_0, t). \tag{1}$$

The velocity $\mathbf{v}_t$ at time $t$ can be obtained through either analytical or numerical differentiation. Expression in Eq. 1 corresponds to the 4D-spacetime model concept realized in the proof-of-concept GICnet model; the latter, in contrast to MDtrajNet, does not include any information about the element types, limiting its transferability.

Due to the chaotic nature of many-body dynamics and practical limitations such as model complexity and data availability, it is generally infeasible to approximate the equations of motion accurately across the entire time axis. As such, the propagation time $t$ must be constrained. In the case of GICnet, a cutoff time $t_c$ (~10 fs, typically) was used, limiting both training and prediction to the interval $t \in [0, t_c]$. This approach proved effective across various organic molecules. Although alternative strategies are possible, we adopt this time-restriction scheme in the present work as well.

Like most machine learning models, MDtrajNet supports batch prediction on multiple times and input configurations simultaneously, enabling highly-parallelized, direct generation of one or more trajectory segments. Under the aforementioned time constraint, each generated segment cannot exceed the duration $t_c$. Therefore, long-time trajectory propagation must be performed by stitching together multiple segments. Nonetheless, compared to traditional molecular dynamics simulations using small time-step integration, this method drastically reduces the number of iterations required. By leveraging parallel predictions within time intervals, it is possible to achieve orders-of-magnitude improvements in simulation efficiency.

In the implementation, the MDtrajNet architecture is primarily composed of multiple attention blocks. Within each block, the atomic positions, velocities, embedded element-type and time features are each updated by corresponding increments. These increments are added to their respective initial values to yield updated representations, which are then passed to the next attention block, if one exists. Through this iterative process, the input features are progressively refined, and the output from the last block provides the increment for the final structure $\mathbf{R}_t$ (Figure 2, left).

In each attention block, the interactions between an atom and its surrounding neighbors are encoded using relative positions $\mathbf{E_R}$, relative velocities $\mathbf{E_v}$, interatomic distances $\mathbf{D}$, element types, and the propagation time. These encoded features are used to construct the queries ($\mathbf{Q}$), keys ($\mathbf{K}$), and values ($\mathbf{V}$) required for multi-head attention, ultimately producing the output increments for the block (Figure 2, right). All operations within this process are designed to preserve O(3)-equivariance, ensuring that the final updated outputs are consistent with the full E(3) symmetry of Euclidean space.

Importantly, in this design, velocities are predicted as derivatives of the structural coordinates. Consequently, both translational shifts of the input structure (i.e., system-wide translations) and of the velocities (i.e., changes in inertial reference frames) do not affect the predicted velocities. The former is a desirable property, while the latter can be trivially corrected using a Galilean transformation, provided that all training samples are expressed in a consistent center-of-mass reference frame.

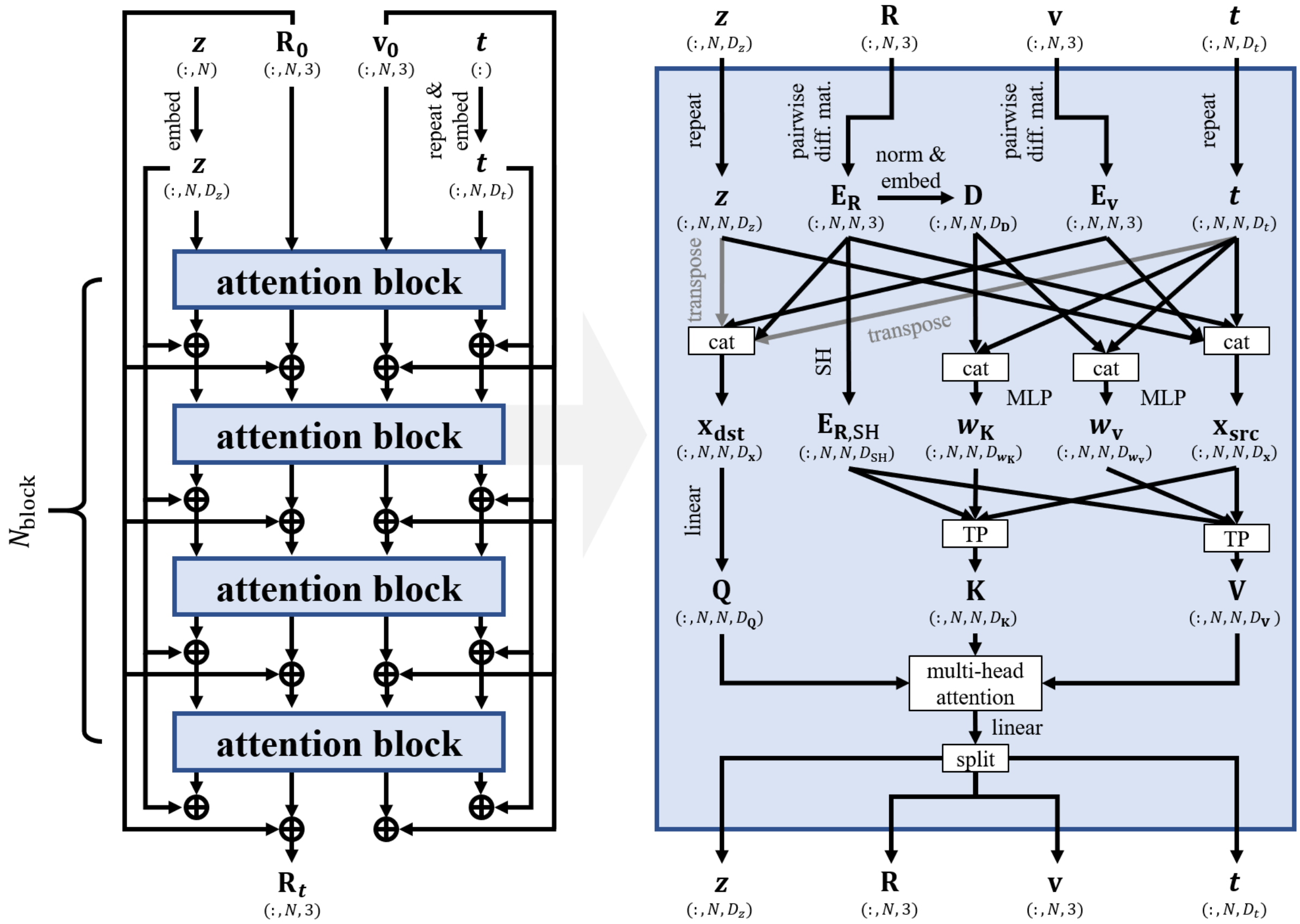

**Figure 2.** The model design of MDtrajNet. The architecture is primarily composed of multiple attention blocks (left). The atomic positions $\mathbf{R}_0$, velocities $\mathbf{v}_0$, embedded element-type $\boldsymbol{z}$ , and embedded time $t$ are each updated by attention blocks in sequence to predict the final structure $\mathbf{R}_t$. In each attention block (right), the interactions between an atom and its surrounding neighbors are encoded using relative positions $\mathbf{E_R}$, relative velocities $\mathbf{E_v}$, interatomic distances $\mathbf{D}$, element types, and the propagation time. These encoded features are used to construct the queries ($\mathbf{Q}$), keys ($\mathbf{K}$), and values ($\mathbf{V}$) required for multi-head attention, ultimately producing the output increments for the block.

Noteworthy, the new MDtrajNet architecture overall enables more accurate learning for a single molecule than our previous GICnet architecture, specifically designed for such cases (see SI).

***Makeup and benchmark of the foundational model MDtrajNet-1***

With the transferable model architecture in hand, we have created the first foundational 4D-spacetime model, MDtrajNet-1. The foundational models need diverse data. For this purpose, we have generated a training dataset, using the uniform settings for the MD trajectory propagations. This dataset comprises one million time steps drawn from the 1-ps-long MD trajectories in the NVE ensemble propagated with 0.1 fs time step starting from 4378 initial conditions corresponding to 173 different molecular systems with two to nine atoms from the ANI-1x dataset[41] (see Methods for details). For the generation of the reference MD data, we have used an accurate, universal MLIP ANI-1ccx, targeting the gold-standard coupled-cluster (CCSD(T)/CBS) accuracy for the same dataset of molecules. CCSD(T)/CBS is a higher-level approach than most commonly used density functional theory (DFT) approaches, i.e., our reference MD trajectories are supposedly of higher quality than typical *ab initio* MD trajectories. This is important as it is known that the quality of MD is often strongly influenced by the quality of the potential rather than by a particular MD algorithm[42-44].

173 training molecular systems is a relatively small number compared to the thousands of unique systems typically used for training the universal MLIPs, e.g., ANI-1ccx was trained on data set with ca. 57k unique systems. Hence, the current generation of MDtrajNet-1 has been trained on a relatively small subset of the chemical space due to the hardware limitations. Also, on average, the training data contains fewer than 10k data points per molecule, which is relatively sparse compared to, e.g., the GICnet model typically trained on ~1 million points for a single molecule. Hence, our foundational models can be considered a lower bound of what is expected for the future models trained on more data. We have trained the models using the 10-fs time cutoff. To improve the robustness of the model, MDtrajNet-1 consists of the ensemble of four models, MDtrajNet-1 (m1–4), each of which can be used separately for faster predictions and fine-tuning. The latter might be useful in resource-constrained scenarios as each model in the ensemble is rather big with ca. 3 million (3 093 568) parameters.

In addition to the training dataset, we have also created an independent test set, containing the seen and unseen molecules, and with energies distributions wider than in the training data making it more challenging (Figure S12). This test dataset consists of 3722 1-ps trajectories of 200 systems (173 seen and 27 unseen) with less than 10 atoms drawn from the ANI-1x. None of the initial conditions in the test set were included in the training set. Inclusion of the seen molecules, for which MD is started from initial conditions different from those in the training set allows us to probe the model's ability to generalize to the unknown phase spaces of the seen molecules. Inclusion of the unseen molecules probes the model's ability to generalize to the new regions of chemical spaces; for this dataset.

We evaluate the MDtrajNet-1 performance in predicting the nuclear positions as a function of time for trajectory segments within a 10-fs cutoff time as the root-mean-squared error (RMSE) with respect to the reference MD trajectory started with the same initial conditions. For the test dataset, MDtrajNet-1 demonstrates its ability to generalize learned knowledge across chemical space by achieving RMSEs below 0.1 Å compared to the reference trajectory and averaged $R^2$ close to 1 for each size of the molecular system, with some outliers possessing unreasonable configurations of error exceeding 1 Å (more on that in the next subsection). Notably, MDtrajNet-1 achieves consistently good performance for molecules of smaller size (fewer than 7 atoms). As the number of atoms increases, accuracy generally decreases, highlighting the challenge of accurately predicting trajectories for more complex molecules and different distributions under limited training data.

It is important to put such performance in perspective, by comparing to established MLIPs. To this end, we trained a reconstructed ANI – a MLIP of ANI-type fitted on the training data of MDtrajNet-1. We also tested 2 sub-models from the reference ANI-1ccx potential, which is an ensemble of eight MLIPs trained on different subsets of a much bigger training set (>50k unique species and 4.5M off-equilibrium configurations). Impressively, MDtrajNet-1 surpasses reconstructed ANI and exhibits performance comparable to the sub-models of the reference universal MLIP ANI-1ccx (Figures S10 and S11).

***Performance of the foundational MDtrajNet-1 model in long-time simulations across chemical space***

A practically useful application of the foundational MDtrajNet-1 model is the long-time MD simulations of diverse compounds. Hence, we explore how this model performs for 200 systems of the test data set by propagating 3722 of 10-ps-long trajectories with MDtrajNet-1, i.e., far longer than the trajectory segments used in the training. To perform such simulations, the model makes predictions within the time cutoff, and the last time step is used as the initial conditions to make predictions again within the time cutoff – this is repeated until the target simulation time is achieved (see Methods). Despite the limitation of the time cutoff, the generation of the MD trajectories with MDtrajNet-1 is orders of magnitude faster than with MLIPs, as we will elaborate later.

Encouragingly, MDtrajNet-1 yields stable long-time trajectories for 97.9% unseen initial conditions in the test dataset with seen molecules (Figure 3c), and for 95.4% of the trajectories of the 27 unseen systems (Figure 3a). The MDtrajNet failures are in the 2–5% cases, mainly for the molecules that have little to no representation in the training data (Figure 3c). This failure rate is likely due to a relatively small coverage of the chemical space in the training test of MDtrajNet-1, i.e., compared to the MLIPs ANI-1ccx and ANI-1x, which were trained on ~330x larger number of compounds. Even then, these MLIPs have very well-known failures in MD, such as unphysical dissociations.[45] The failure rates can be decreased by refining the data sets in the regions relevant for the application at hand, e.g., via active learning as is done in MLIPs.

To judge the quality of the trajectories, we compare the similarity of the corresponding power spectra derived from the MDtrajNet-generated and reference trajectories as measured by the Pearson correlation coefficient (using the standard definition[46] for spectra). Such a comparison is rather challenging, and perfect similarities are not observed even for the MLIPs trained on lots of data for a single molecule, which can have similarities in power spectra as low as 0.72.[15] Here, we have a more challenging task of evaluating the models' performance for many different molecules.

Hence, to put the performance of the MDtrajNet-1 across chemical space in perspective, we compare it to the reconstructed ANI (a MLIP trained on the same training data as used for MDtrajNet-1), both of which attempt to approximate CCSD(T)/CBS-level dynamics. We also compare to another universal MLIP, ANI-1x,[41] pre-trained on the DFT (ωB97X/def2-TZVPP). Hence, the performance of the ANI-1x trajectories is approaching that of AIMD propagated with the commonly used DFT. Note also that ANI-1ccx was derived from ANI-1x via transfer learning using a limited number of CCSD(T)/CBS training points. As a result, the two models share a substantial portion of training data and model parameters.

Impressively, our foundational MDtrajNet-1 model produces less divergent trajectories from the reference ANI-1ccx MD, compared with the reconstructed ANI trained on the MDtrajNet-1 training data (Figure 3a–b). MDtrajNet-1 also deviates as much as the DFT-level AIMD trajectories (Figure S13). Among the 3644 trajectories that were successfully simulated by both MDtrajNet-1 and the reconstructed MLIP, MDtrajNet-1 achieved higher spectral similarity in nearly 56% (2032) of cases (Figure 3d). Even for molecules with lower similarity scores, the model can accurately predict vibrational peak frequencies (Figure 3e–f). These results suggest that the errors of MDtrajNet-1 in long-time simulations are small enough, and this foundational model can successfully reproduce vibrational behavior across diverse compounds. This is an important observation, meaning that directly learning dynamics has the potential to produce more robust MD results than learning forces.

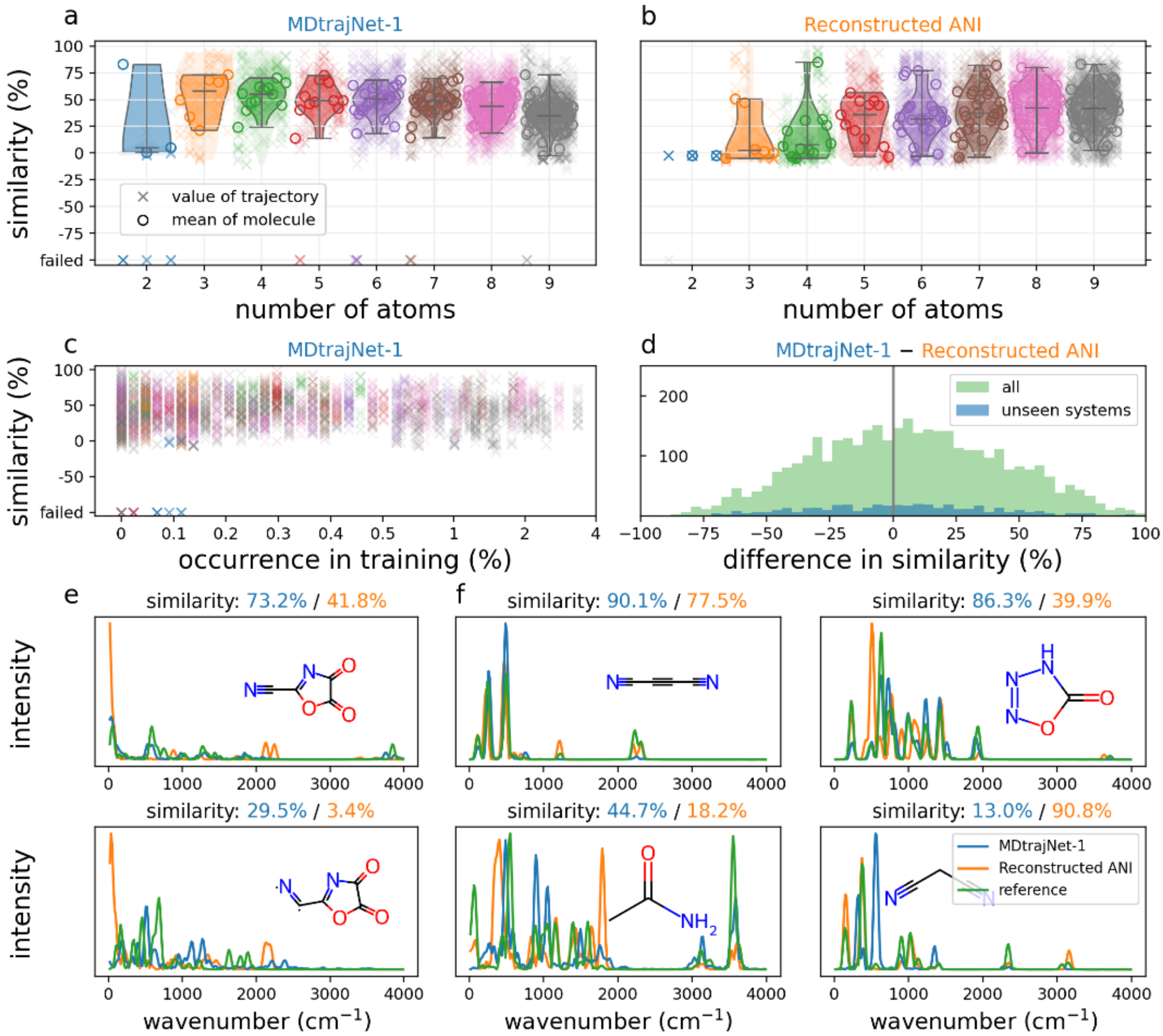

**Figure 3.** a, b) The similarity of power spectra for the 10-ps-long test trajectories generated with a) the foundational MDtrajNet-1 and b) the classical MD propagated with 0.05 fs time step using the reconstructed ANI that was trained on the same dataset as used for MDtrajNet-1. Each faint cross in the left plot represents the value for an individual test trajectory, while a circle represents the mean value of a tested molecule. Their distributions are also shown by the violin plots, whose ticks indicate the extreme and median values. Results of different systems are shown with identical horizontal offsets, which are further grouped by the number of atoms in the system. c) The spectral similarities of the test set versus the occurrence of the configurations of the corresponding test system in the training data

(these training configurations are not used in any of the plots in this figure). d) The distribution of the differences in similarity in the stable trajectories of all and unseen molecules in the test data set. The difference is obtained by subtracting the similarity scores of ANI-1x from those of MDtrajNet-1 wrt the reference ANI-1ccx trajectories. e, f) Representative spectra with varying similarities for unseen (e) and seen (f) molecules. Note that initial conditions for seen molecules were not included in the training set.

The biggest advantage of the 4D spacetime model compared to the traditional MD approaches is the efficiency, which comes from the force-free propagation and a significantly reduced number of iterative steps. Compared with the commonly employed acceleration of MD with ML used only for the faster calculation of the forces, the MDtrajNet model can achieve an efficiency boost of 2 orders of magnitude, as shown in Figure 4a when compared with the finest temporal resolution (0.05 fs time step) tested.

Here, we use the term *temporal resolution* for the target time step, used to sample configurations from the trajectories to post-process in applications such as power or vibrational spectra simulations. It is distinct from the time step used in traditional MD propagation, i.e., the minimum time interval in an MD trajectory. The temporal resolution is oriented towards the application and thus depends on it, while the propagation time step dictates the trajectory quality in traditional MD (Figure 4d), where the time step generally has to be much smaller than the temporal resolution needed. For example, to generate a vibrational spectrum that covers 0 to 4000 $cm^{-1}$ wavenumbers, a temporal resolution finer than 4.17 fs is enough to satisfy the criterion set by the Nyquist–Shannon sampling theorem[47, 48]. However, a time step close to that criterion might lead to an unstable trajectory in traditional MD (Figure 4a, orange crosses, and Figure 4d ) due to numerical integration errors; therefore, a smaller one has to be used. Hence, oftentimes, smaller time steps are used for propagation, but only some of them are needed in an application. For example, if the time step of 0.5 fs is used, but the trajectory is saved every 4 fs, it effectively leads to the 'waste' of ca. 90% of calculations, which are only needed to ensure the correct propagation with the traditional MD algorithms.

In contrast, the foundational MDtrajNet-1 model always gives its best quality predictions within the time segment, regardless of the time step (Figure 4d), so that we can choose the temporal resolution in the MDtrajNet-generated trajectories closer to the minimum requirement in the application. Taking this into account, the MDtrajNet model can be used to only make predictions at the required time steps, e.g., 4 fs in the above example, further boosting the efficiency of the MD in real applications and lowering resource consumption by eliminating the need for intermediate, unused time steps.

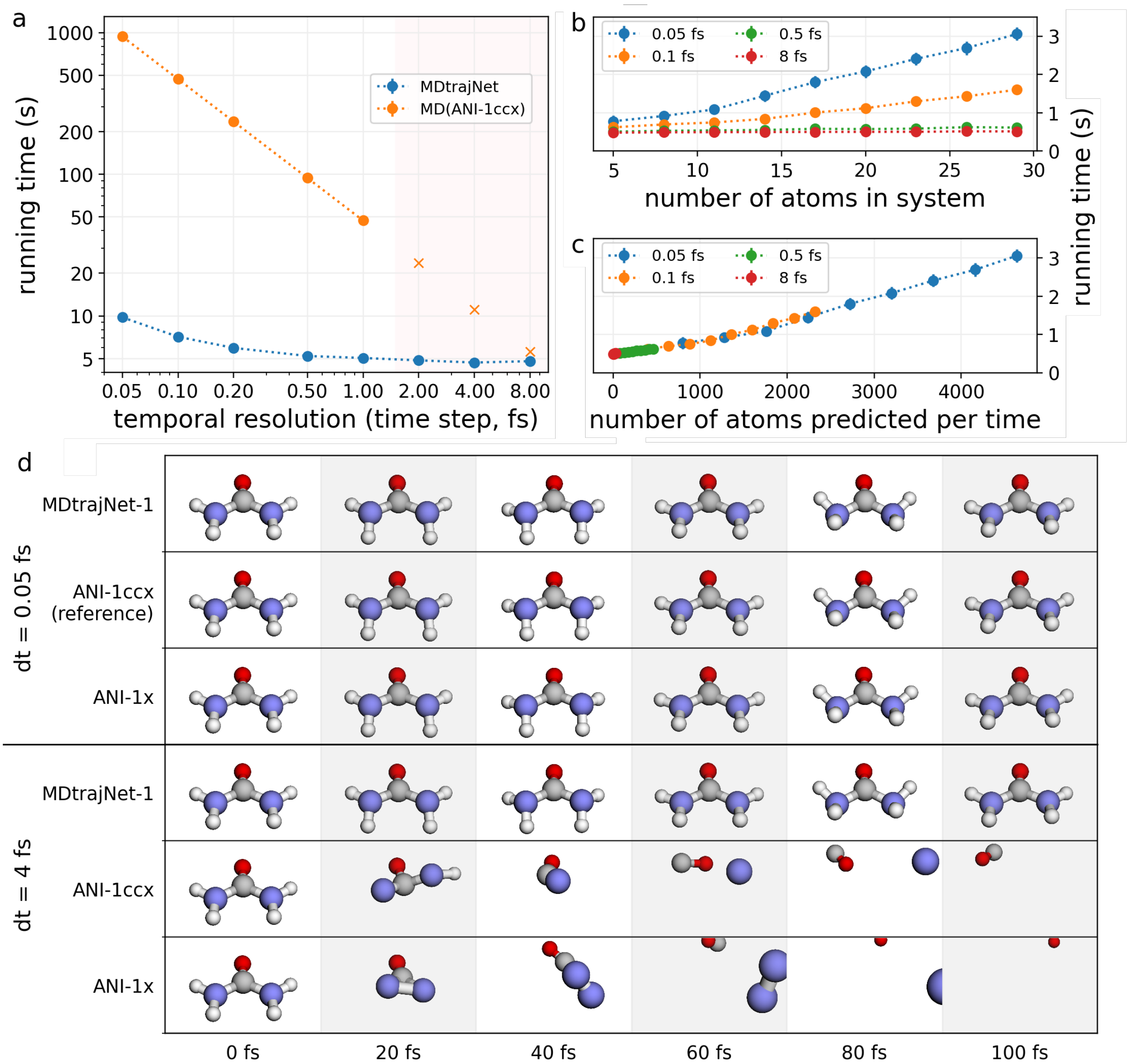

**Figure 4.** a) The wall-clock (running) times used to propagate a 1-ps trajectory for an ethanol molecule with different time steps. The results of a single MDtrajNet-1 submodel (blue) were simulated with a time segment of 8 fs, and different settings of time steps do not affect the quality of each frame. For the traditional MD with ANI-1ccx potential (orange), larger time steps deteriorate the quality of integration: the temporal resolution is equivalent to the time step, and time steps larger than 1 fs led to the broken molecular geometries (crosses). Each running time is reported as an average of five repeats. b) The running times in seconds for different system sizes using a single MDtrajNet-1 submodel. Four different temporal resolutions (time steps) were used to propagate a 100-fs trajectory with 8-fs segments. c) The same result in (b) plotted using the number of atoms predicted per time in a segment as the x-axis. All timings are reported for one Nvidia RTX 4090 GPU with 16 cores of Intel Xeon Gold 6526Y@3.9GHz CPUs. d) Snapshots from the MD trajectories of the urea molecule generated with the foundational MDtrajNet-1 and propagated with MLIPs ANI-1ccx[10] and ANI-1x[41] using two different time steps (0.05 and 4 fs). The quality of trajectories propagated with the conventional MD with MLIPs strongly depends on the chosen time step – the limitation lifted by the MDtrajNet-1.

Another important issue is the scaling of the computational time with respect to the system size. The propagation efficiency of MDtrajNet has an approximately linear dependence between the time required to simulate trajectories of the same length and the number of atoms in the system (Figure 4b,c). This linearity arises from the model architecture, which only considers the local environment around each central atom and masks interactions between atom pairs beyond a certain radial cutoff. Consequently, for larger systems, the number of atomic interactions to be computed increases linearly with the number of atoms, assuming a constant local atomic density per atom, rather than exhibiting quadratic scaling. Furthermore, although the inference efficiency differs significantly across segments with varying temporal resolutions (Figure 4b), the prediction efficiency with respect to atomic position remains roughly consistent (Figure 4c). This suggests an inverse relationship between system size and temporal resolution under a given computational budget and inference efficiency. Hence, MDtrajNet-1 can push the boundaries of the MD simulations also with respect to the system size.

***Transferability and fine-tuning***

The foundational MDtrajNet-1 model has been trained on relatively small systems with nine or fewer atoms, while the systems of practical interest can be much bigger. Hence, we explore the applicability of the MDtrajNet model architecture to systems with more than nine atoms. Here, we leverage the foundational MDtrajNet-1 model by fine-tuning its pre-trained parameters for systems with more than nine atoms. Our tests show that fine-tuning is more efficient than training the model from scratch, indicating that the foundational model already possesses some predictive power for larger systems (compare panels a and d, Figure 5).

As a test system, we take alanine dipeptide, a 22-atom system, which is a prototypical model system of peptides commonly used to evaluate the performance of force fields in simulating protein conformational changes. We use this system to explore our model performance by examining the dipeptide's conformational space in terms of φ and ψ dihedral angles. that are critical to polypeptide and protein conformations (Figure 5c).

We construct the training set from very short MD trajectories (32 ps in total) propagated with the reference ANI-1ccx potential (see Methods). One submodel picked from the MDtrajNet-1 (m1–4) foundational model ensemble is then fine-tuned on this data set and used to generate a long, 1-ns, MD trajectory. Remarkably, the fine-tuned model is capable of capturing the conformational space in good agreement with the reference 1-ns trajectory propagated with ANI-1ccx (Figure 5a,b).

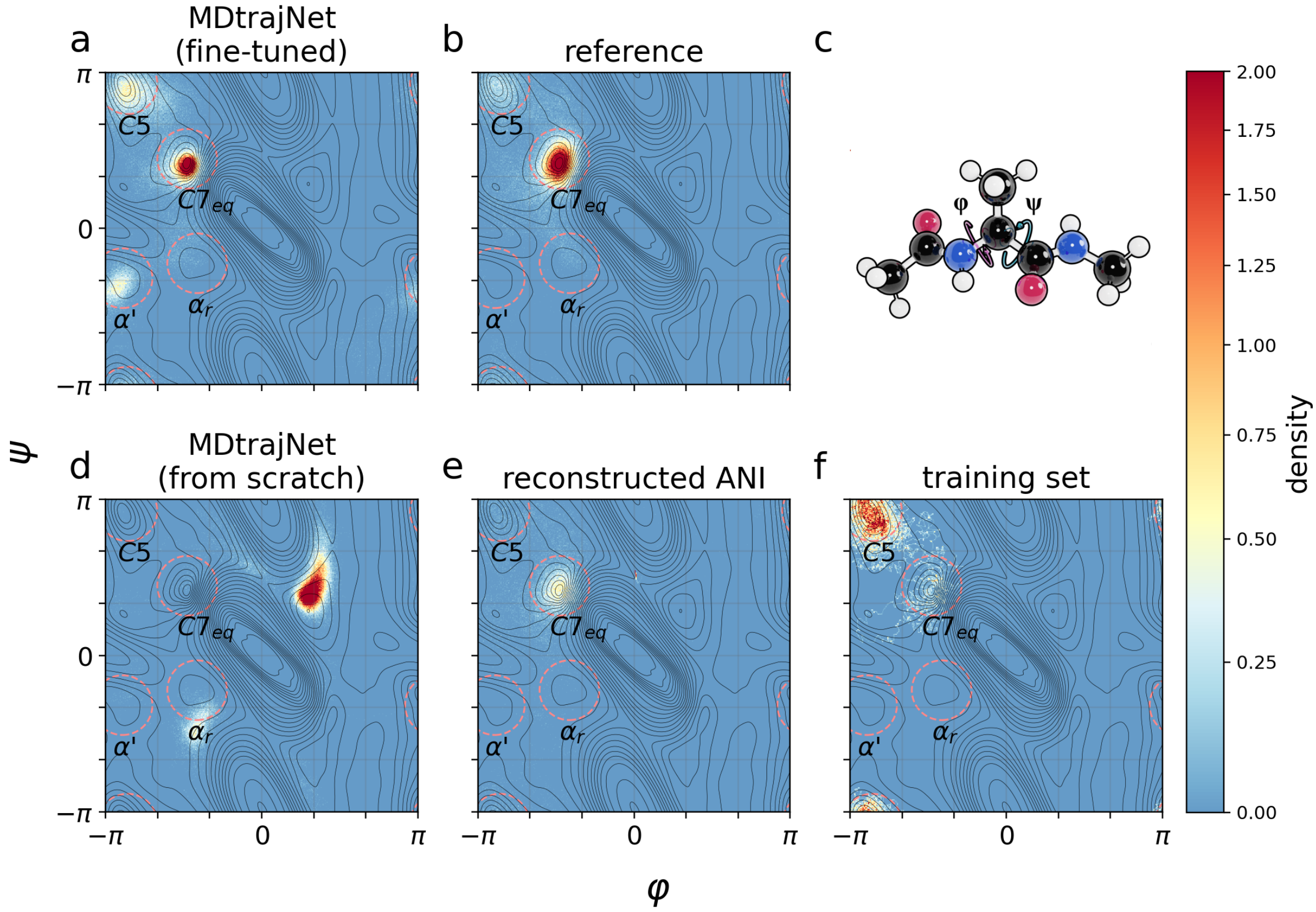

**Figure 5.** The Ramachandran plots of the alanine dipeptide molecule (c) from MDtrajNet and MLIP potential simulations. The geometry distributions from a) the 1-ns trajectory from the fine-tuned MDtrajNet model, b) 1-ns MD trajectory propagated with the reference method ANI-1ccx from the same initial conditions, d) 1-ns MD trajectory propagated with the MDtrajNet model training from scratch on the alanine dipeptide dataset, e) 1-ns MD trajectory with an MLIP (ANI) potential model trained on the alanine dipeptide dataset, f) the geometry distribution in the training trajectories (32 ps in total) of the MDtrajNet model. Densities are normalized based on their range by shifting and taking a logarithm. A warmer color represents a higher value, and *vice versa.* All trajectories are propagated from the same initial conditions.

The MDtrajNet model reproduces well the population distribution between C5 and $C7_{eq}$ (using notations from Ref. [49]) relative to that at the reference potential and visits the $\alpha_r$ or α' regions. This is rather impressive given that the training data (Figure 5f) mostly covers the C5 region, with minimal samples near $C7_{eq}$ and none in $\alpha_r$ (the right-hand alpha helix) or α' (representing meta-stable conformations). These deficiencies in the training data are well compensated for by the generalization ability of the pre-trained parameters in the foundational MDtrajNet-1 model.

This result is even more striking when comparing the fine-tuned MDtrajNet-1 performance to that of training from scratch more established MLIPs. Here we train from scratch the MLIP of ANI type on the same data as used for fine-tuning pre-trained MDtrajNet-1 and use this MLIP model to propagate the 1-ns MD. The MLIP trained from scratch does not reflect the correct distribution between C5 and $C7_{eq}$ (Figure 5e). Furthermore, it falls into an abnormal potential well to the right of $C7_{eq}$ after ~300 ps. This demonstrates the challenge of training on poorly distributed data without a pre-trained model and underscores the importance of our MDtrajNet-1 model's transferability.

Overall, the above experiments highlight the transferability of the MDtrajNet architecture due to its atom-centered design, enabling effective generalization to larger and previously unseen molecules. The model not only learns from single molecular systems but also shows an ability to generalize and integrate knowledge across different molecular systems when handling trajectory data.

***NVT ensemble***

In the previous sections, all tests on different molecular systems used trajectories generated under the NVE ensemble. However, this is not a strict requirement for the MDtrajNet architecture. Thus, we also present a preliminary demonstration of the MDtrajNet architecture's predictive ability under statistical ensembles other than NVE, using the NVT ensemble to generate reference trajectories.

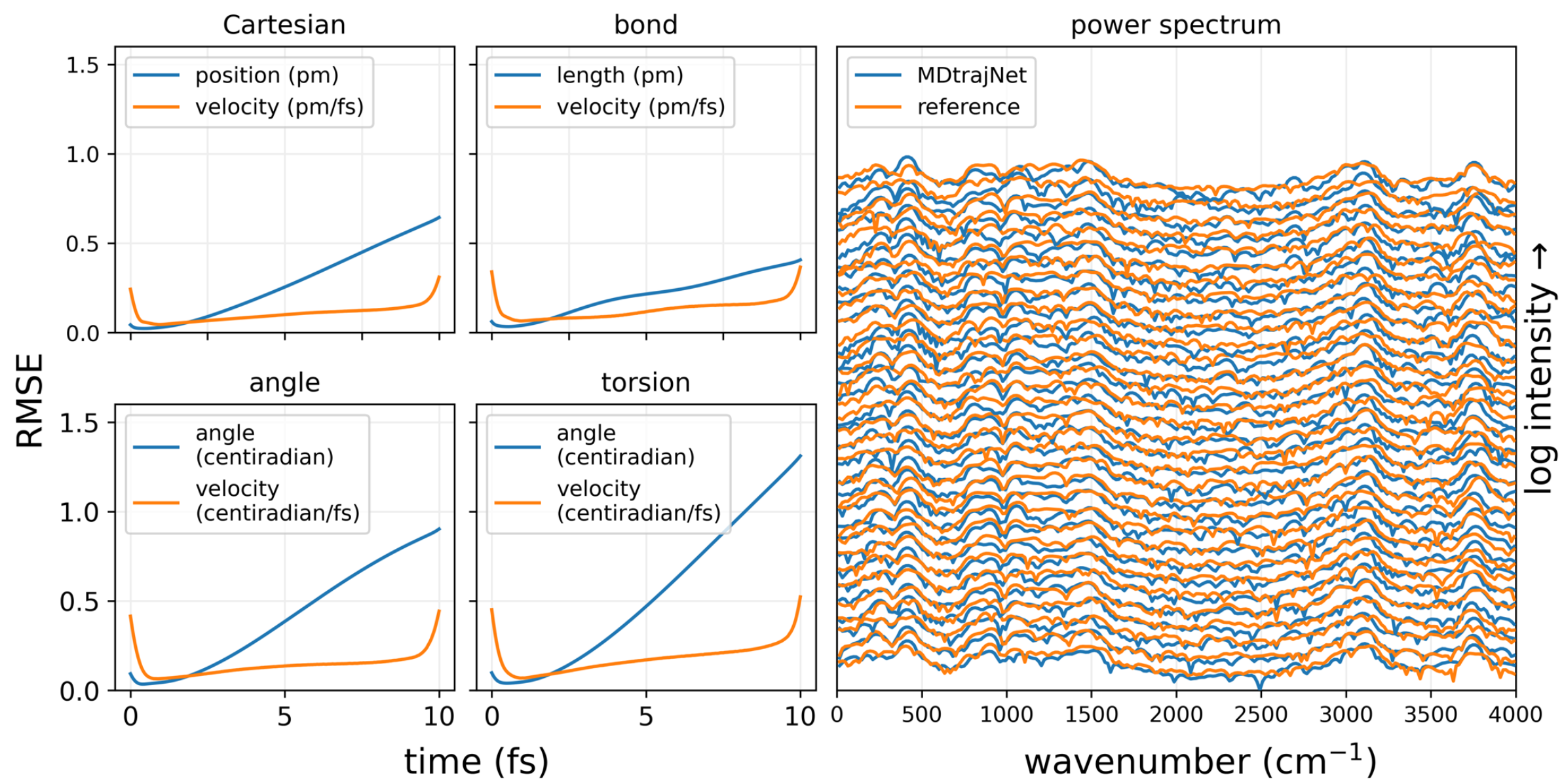

**Figure 6.** The prediction errors within the time cutoff and the power spectra (multiple test trajectories) predicted from the model based on the MDtrajNet architecture and trained on the NVT data for the ethanol molecule.

The performance of the ethanol NVT model on the test set is shown in Table S1. Compared to the model trained and tested on NVE trajectories, the performance under NVT appears slightly inferior. This may be attributed to the additional degrees of freedom introduced by the thermostat parameters in the Nosé–Hoover method, which effectively increases the complexity of the system and the training difficulty. Nevertheless, the ethanol model still achieves very low prediction errors within the time cutoff, as illustrated in Figure 6.

For evaluating long-time NVT trajectory predictions, the power spectra are also shown in Figure 6. The results demonstrate that the ethanol model can still accurately reproduce molecular vibrational modes in long-time NVT simulations.

These findings indicate that the MDtrajNet architecture is not restricted to NVE ensemble trajectory predictions and can be successfully applied to other ensembles as well. However, a notable limitation is that the current design of the MDtrajNet model does not explicitly incorporate additional thermodynamic parameters (e.g., temperature). Therefore, currently, for different ensembles that require parameters such as temperature or pressure, the model needs to be separately trained to account for those variables, but this limitation can be solved in future versions of MDtrajNet.

***Periodic systems***

Due to its atom-centred local representation, the MDtrajNet architecture can also be employed in simulations with periodic boundary conditions. Here, we present two application cases to validate this capability.

The first case employs a 2×2×2 supercell of diamond containing 64 carbon atoms (Figure 7a), for which we have propagated a reference trajectory in the NVE ensemble with ANI-1xnr potential[50] using initial conditions equilibrated at 300 K. The training on this data converged after just 6 epochs, achieving an accuracy of 0.0028 Å on the validation set. The resulting diamond model generates stable dynamics of this periodic system over longer timescales and accurately reproduces both the peak positions and widths of the radial distribution function (RDF) of carbon atoms in the system compared to the reference 1-ps trajectories (Figure 7b). This demonstrates the applicability of the MDtrajNet architecture to periodic systems.

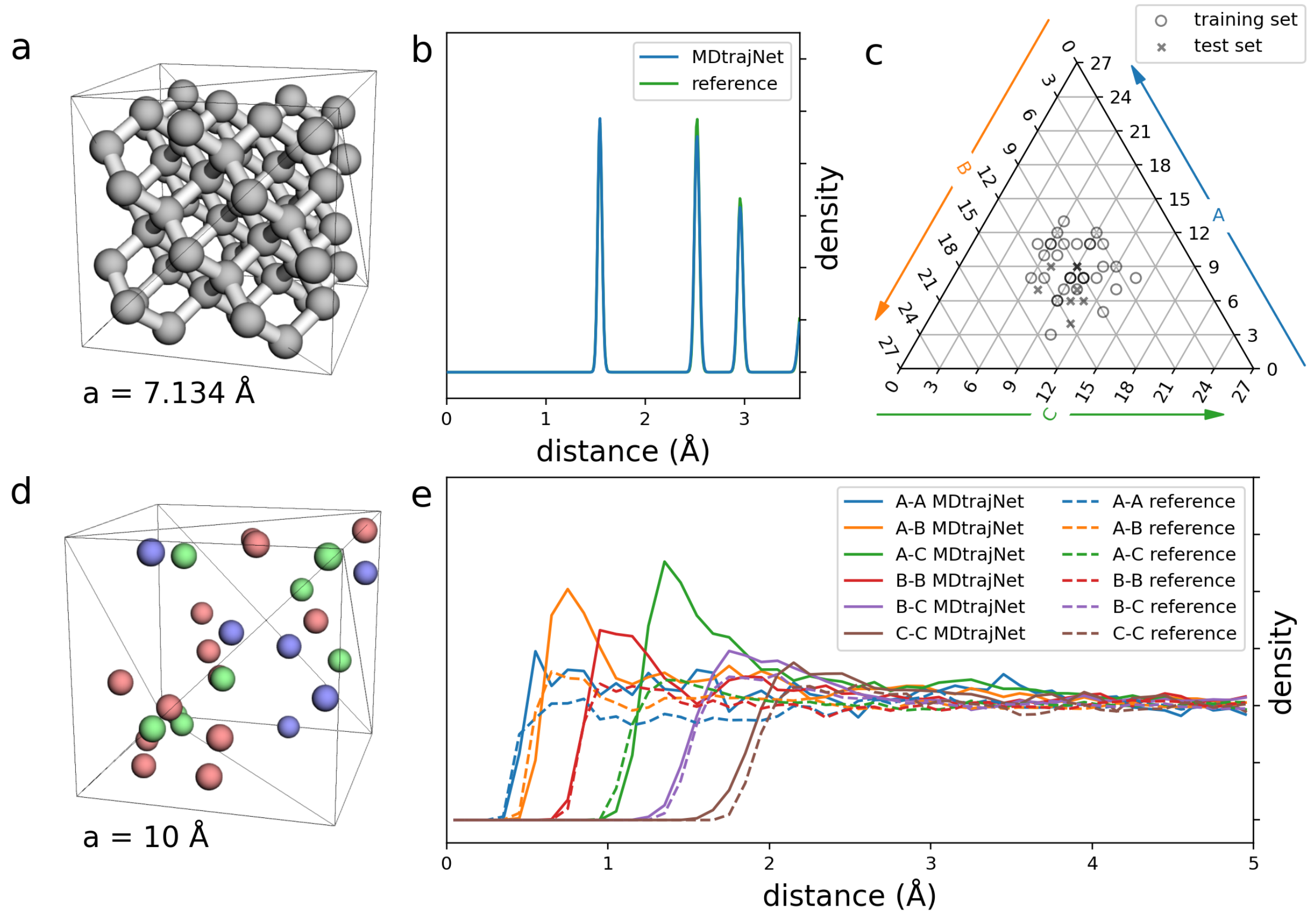

**Figure 7. Application of the MDtrajNet architecture to simulations with periodic boundary conditions.** a) The 3D representation of the diamond system in a periodic cell. b) The radial distribution of the distances between carbon atoms in the diamond system. Distances were sampled from 64 1-ps trajectories propagated with MDtrajNet-based model and the reference method (traditional MD with ANI-1xnr potential[50]). c) The compositions of particle types in the training and test set for the systems of 27 virtual particles in 3 different types (denoted as A, B, and C). d) The 3D representation of a system of virtual particles in a periodic cell. e) The radial distribution of different particle pairs in eight 10-ps test trajectories of the systems of virtual particles with the Lennard-Jones potential. The distributions from the MDtrajNet-generated and reference trajectories are shown with solid and dashed lines, respectively.

To further test this capability with a more flexible scenario, the second application case employs 27 virtual particles of three different types (denoted as A, B, and C) confined within a periodic cubic box of 1 nm per side as shown in Figure 7d. The composition ratios of these particle types are varied across different simulations (Figure 7c). Moreover, the interactions between particles are described using the Lennard-Jones potential[48], which is fundamentally different from the previously used ones. The parameters used for each particle pair are provided in Table S2.

The radial distributions of inter-particle distances obtained from the model based on MDtrajNet architecture and trained on the Lennard-Jones data are compared with those from the reference simulations, as shown in Figure 7e. The results indicate that the MDtrajNet-based model reproduces the overall shape of RDFs and captures the relative positions of the distribution peaks corresponding to different interaction types, although densities are sometimes overestimated.

These results on a periodic system with Lennard-Jones interactions demonstrate that the MDtrajNet architecture is applicable to diverse interaction types and periodic systems, further confirming the versatility and generality of its design.

**Conclusions**

In this work, we present a novel neural network architecture MDtrajNet for directly generating molecular dynamics trajectories, circumventing the need for the iterative solving of the Newtonian equations of motion and force calculations. We also present a foundational MDtrajNet-1 model, based on this architecture and pre-trained on a big data set. MDtrajNet-1 is transferable to a wide range of molecular systems, allowing efficient generation of the trajectories hundred times faster and with higher accuracy than established MLIP trained on the same data. The architecture and foundation models represent a significant milestone in the development of the emerging class of 4D-spacetime AI models learning the time-evolution of atomistic systems.[15]

The key to the model's transferability is incorporating equivariant neural networks in its architecture, which enables the model to maintain the physical requirement of E(3)-equivariance when directly using Cartesian coordinates. Additionally, leveraging the transformer architecture enables the model to hierarchically assess the influence of neighboring atoms centered around each atomic environment. This allows MDtrajNet to predict molecular structures at future time frames while ensuring invariance to the ordering of atoms in the input.

These architectural innovations empower a single MDtrajNet model to make accurate predictions across systems with diverse elemental compositions and varying numbers of atoms, overcoming a key limitation of the previous 4D-spacetime GICnet model,[15] which lacked such generality. Moreover, MDtrajNet achieves enhanced predictive accuracy and stability.

Performance evaluations on both short- and long-timescale simulations demonstrate the strong generalization ability of the foundational MDtrajNet-1 model, which was trained on hundreds of molecular systems. Even with limited training data, it delivers accurate predictions for a wide variety of molecular systems, including those that are not included in the training data. This reflects its high data efficiency and capacity to learn robust representations of dynamical behavior. MDtrajNet's better performance than that of traditional MLIP trained on the same data indicates that directly learning dynamics has the potential to produce more robust MD results than learning forces.

Furthermore, fine-tuning of MDtrajNet-1 can be used to substantially extend its extrapolation ability to entirely new systems with atom counts far beyond those seen during training. This illustrates the model's scalability and adaptability. The underlying MDtrajNet architecture also shows excellent applicability to different statistical ensembles and periodic systems, further highlighting its versatility.

Taken together, the accuracy, generality, and efficiency demonstrated by MDtrajNet underscore the immense potential of the 4D-spacetime framework as a new paradigm for molecular dynamics. This approach promises to benefit a wide array of applications requiring large-scale, long-timescale trajectory simulations. As with all data-driven methods, we expect substantial future improvements with the increased availability of extensive, high-quality training datasets, advancement in hardware, and further architectural improvements.

## Methods

### *Further details on neural network architecture*

*The following description is given for a single, non-batched data entry for clarity.*

As shown in Figure 2, the MDtrajNet model requires 4 inputs: atom types $\boldsymbol{z}$, initial coordinates $\mathbf{R}_0$ , initial velocities $\mathbf{v}_0$, and a time duration $t$ (Figure 2). Before passing them into attention blocks, the atomic species $\boldsymbol{z}$ are embedded to $\mathbf{z}$ via a trainable embedding layer with an embedding dimension of $D_\mathrm{z}$ , and the time duration $t$ is encoded to $\mathbf{t}$ by the *soft_one_hot_linspace* function implemented in the *e3nn* Python module[51] with a specified cutoff $t_\mathrm{c}$ and embedding length $D_t$.

The model then adopts a transformer-based architecture that consists of $N_\mathrm{block}$ equivariant attention blocks, which is modified from SE(3)-transformers[40] to incorporate extra input, i.e., velocities, embedded time duration, and other relevant features (e.g., related to PBC and cell parameters). Those blocks are sequentially joined and each of them transforms the four inputs ($\mathbf{z}$, $\mathbf{R}$, $\mathbf{v}$, $\mathbf{t}$) to corresponding increments ($\mathbf{\Delta z}$, $\mathbf{\Delta R}$, $\mathbf{\Delta v}$, $\mathbf{\Delta t}$) for updating the original inputs. The final prediction of the model is obtained by adding the $\mathbf{\Delta R}$ of the last block to $\mathbf{R}_0$.

Inside each block, transformations are performed before calculating the attention. First, each input is be transformed to a 3-order tensor of shape ($N_\mathrm{atom}$, $N_\mathrm{atom}$, 3), where $N_\mathrm{atom}$ is the number of atoms in the system. For $\mathbf{z}$ and $\mathbf{t}$, this is done by simply repeating the matrix in the second last axis of the final tensor, while for $\mathbf{R}$ and $\mathbf{v}$, the edge vector (or relative positions) for each ordered pair (source → destination) is computed from them to give edge tensors $\mathbf{E_R}$ and $\mathbf{E_v}$. For example, each element in $\mathbf{E_R}$ is defined as:

$$\mathbf{E}_{\mathbf{R},ijk} = \mathbf{R}_{jk} - \mathbf{R}_{ik}. \tag{2}$$

Both of the two tensors are in the shape of ($N_\mathrm{atom}$, $N_\mathrm{atom}$, 3) as desired. Then the interatomic distances can be calculated by taking the 2-norm of $\mathbf{E_R}$, which are further embedded to the tensor $\mathbf{D}$ using *soft_one_hot_linspace* with a radial cutoff $R_\mathrm{c}$ and an embedding length $D_\mathrm{d}$. Meanwhile, the spherical harmonics representation of $\mathbf{E_R}$ (denoted as $\mathbf{E}_\mathrm{SH}$) is derived with a maximum order $l_\mathrm{max}$.

Two feature tensors $\mathbf{X}_\mathrm{src}$ and $\mathbf{X}_\mathrm{dst}$ that corresponds to the two ends of atom pairs can be defined by concatenating $\mathbf{z}$, $\mathbf{E_R}$, $\mathbf{E_v}$ and $\mathbf{t}$ in the last axis:

$$\mathbf{X}_\mathrm{src} = \left[\mathbf{z}; -\frac{\mathbf{1}}{\mathbf{2}}\mathbf{E_R}; -\frac{1}{2}\mathbf{E_v}; \mathbf{t}\right]_3, \tag{3}$$

$$\mathbf{X}_{\mathrm{dst}} = \left[\mathbf{z}^{\mathrm{T}}; \frac{\mathbf{1}}{\mathbf{2}}\mathbf{E}_{\mathbf{R}}; \frac{1}{2}\mathbf{E}_{\mathbf{v}}; \mathbf{t}^{\mathrm{T}}\right]_{3}. \tag{4}$$

The tensors $\mathbf{X}_{\mathrm{src}}$ and $\mathbf{X}_{\mathrm{dst}}$ are used in calculating the query, key, value ($\mathbf{Q}$, $\mathbf{K}$, $\mathbf{V}$) for the multi-head attention. The $\mathbf{Q}$ can simply be obtained via an O(3)-equivariant linear transform $\mathbf{W}_{\mathrm{q}}$ applied to $\mathbf{X}_{\mathrm{dst}}$, while $\mathbf{K}$ and $\mathbf{V}$ are obtained by calculating the fully-connected tensor product of $\mathbf{E}_{\mathrm{SH}}$ and $\mathbf{X}_{\mathrm{src}}$, where the weights of different paths are controlled by both $\mathbf{D}$ and $\mathbf{t}$, through corresponding feed-forward neural networks.

Then the multi-head attention mechanism is executed using the scalar tensor product of the irreducible representations instead of the commonly used dot product to maintain the equivariance. Finally, the increments are given by splitting output of multi-head attention in the same but reversed way of forming $\mathbf{X}_{\mathrm{src}}$ after an O(3)-equivariant linear transform.

Note that in computing the edge vector of each atom pair, the coordinates of the second atom in adjacent cells will be considered when periodic boundary conditions are enabled. However, only the position that gives the minimum distance will be taken. Thus, a cell size that is at least two times larger than the radial cutoff is required.

**Table 1.** The default values of model parameters used in this work.

| parameter | value | description |
|---|---|---|
| $D_z$ | 16 | dimension of embedded species |
| $D_t$ | 4 | dimension of embedded time |
| $N_{\mathrm{block}}$ | 8 | number of attention blocks |
| $N_{\mathrm{head}}$ | 4 | number of heads |
| $R_{\mathrm{c}}$ | 4 Å | radial cutoff |
| $D_d$ | 16 | dimension of embedded distances |
| $l_{\mathrm{max}}$ | 3 | maximum $l$ for spherical harmonics |
| $I_{\mathrm{q}}$ | 8×0e + 8×1o | irreducible representation for query |
| $I_{\mathrm{k}}$ | 8×0e + 8×1o | irreducible representation for key |
| $I_{\mathrm{v}}$ | 8×0e + 8×1o | irreducible representation for value |
| neurons for key product | [128] | to obtain weights in tensor product calculations |
| neurons for value product | [128] | to obtain weights in tensor product calculations |

***Data transformation***

The data format that MDtrajNet transformer fits consists of: a pair of two points (initial/final) in the phase space sampled from the same trajectory, the time interval between them, and the atomic composition of the system. The atomic composition ($\boldsymbol{z}$), Cartesian coordinates of the initial point ($\mathbf{R}_0$), Cartesian velocities of the initial point ($\mathbf{v}_0$), and the time interval ($t$) are used as the input features. The Cartesian coordinates and velocities of the final point ($\mathbf{R}_t$ and $\mathbf{v}_t$) serve as the labels.

To sample data from a reference trajectory with a given cutoff on the time interval, in this work, we split one long trajectory into short trajectories that have a length of the cutoff. For each short trajectory, all points were paired with the first point to generate the desired data.

***Model training***

*Unless otherwise specified, all simulations are performed with MLatom 3[52] software.*

*General overview*

Before training, two subsets will be randomly sampled from the whole dataset for network training and validation. The size of each set depends on the value specified.

The hyperparameters adopted for the model definition in this work are listed in Table 1. The implemented training of all parameters in MDtrajNet models uses the *AdamW* optimizer and starts with an initial learning rate, which is later scheduled in the training via a *ReduceLROnPlateau* scheduler. In this work, we used a scheme where the learning rate starts from 0.001 and the scheduler halves it whenever the validation loss is not improved for 16 consecutive epochs. The training process terminates right after the learning rate drops below $10^{-6}$. Unlike our previous work[15], the RMSE (root-mean-squared error) on the molecular geometries is used as the sole metric in the loss function. The RMSEs in velocities and potential energies are not included in the backpropagation of the networks due to concerns of efficiency. The forward and backward propagations are executed in mini-batches of a given size. Each batch only contains the data points with a uniform sequence length (or number of atoms). The mini-batches are also used in the validation; only a specified number of batches randomly chosen from the validation set will be validated at the end of an epoch.

*Ethanol models*

To generate the reference trajectories, 33 initial conditions were sampled with the Wigner distribution in the ground vibrational state[53, 54] as described in our previous work[15]. After removing their linear and angular momenta, MD simulations were performed with the resulting initial conditions using the ANI-1ccx potential until the target length of 10 ps was reached for each trajectory. A time step of 0.05 fs was chosen to give a fine time resolution. For each initial condition, two trajectories with NVE and NVT (the Nosé–Hoover thermostat[55, 56], 300 K) ensembles were obtained.

For either the NVE or the NVT MDtrajNet models of ethanol, 30 corresponding reference trajectories were used to generate the dataset with a time cutoff of 10 fs. The training and validation sets contain 500k and 50k points, respectively, along with a batch size of 800 points for both of them. Only 1 batch from the validation set was tested at the end of an epoch.

To test the impact of learning the entire time segment instead of just the end of it, we filtered the NVE dataset by removing all points with time below 8 fs. With the filtered dataset, a model was trained with the setup but with only 100k training points (10k validation), which preserves the amount of data in the 8–10 fs region. We also trained another model with the non-filtered dataset with 100k training points for further comparison on the impact of learning adjacent time frames (Figure S4).

*The foundational MDtrajNet-1 submodels*

The ANI-1x dataset[41] features ~5 M geometries in total for 3114 atomistic systems that consist of 2 to 63 atoms. In this work, we only selected 4378 geometries as initial geometries from the 173 systems containing fewer than 10 atoms for the training and validation dataset. Geometries with high potential energies were removed to filter out highly distorted geometries, thus improving the stability of the following MD simulation and also reducing the complexity of the phase space explored. The percentages of atomic systems are roughly preserved after selection. For testing purposes, we also chose 200 systems with 3722 geometries from the ANI-1x dataset in the same range of atom numbers, which also contains systems not present in the previous selection.

For all selected initial geometries, random velocities sampled from the Maxwell–Boltzmann distribution at 300 K were assigned to them, and the linear and angular momenta were eliminated. Finally, MD simulations were performed in the NVE ensemble using the ANI-1ccx potential, with a time step of 0.1 fs until reaching the maximum propagation time of 1 ps.

The trajectories from 4378 initial conditions were used to generate the dataset with a 10 fs time cutoff. Four models were trained with 1M training points and 800 points per batch. The validation set contained 8192 points, all of which were used for validation at every epoch.

*Fine-tuning – alanine*

We first generated 32 NVE trajectories of 1 ps length using the universal ANI-1ccx potential in conventional molecular dynamics propagation. The initial structures were generated using Open Babel[57], and initial velocities were sampled from a Maxwell–Boltzmann distribution at 300 K, with a time step of 0.05 fs.

Subsequently, we performed fine-tuing using one model from the MDtrajNet-1 ensemble as the source model, fine-tuning it with 10k, 50k, 100k, and 500k randomly sampled data points. As a comparison, models were also trained from scratch using 100k and 500k data points.

The training used the same configurations as in the training MDtrajNet-1 models, except that the batch size and the number of validation batches per epoch were changed to 128 and 4. Training metrics were recorded for each case.

*Fine-tuning– alanine dipeptide*

We generated 64 trajectories of 1 ps each for the alanine dipeptide using the same setup as for alanine. Then, 500K data points were randomly sampled for fine-tuning using one of the models from the MDtrajNet-1 (m1–4) ensemble. Training was done with the same settings as for alanine.

Additionally, all training trajectories (sampled at 1 fs intervals) were used to train an ANI model with randomly initialized weights from scratch, aiming to reconstruct the ANI-1ccx potential energy surface for alanine dipeptide.

*Diamond supercell*

A total number of 64 1-ps NVE trajectories (0.05 fs time step) of the 2×2×2 supercell of diamond were generated using ANI-1xnr potential[50] with initial conditions incubated at 300 K for 200 fs (Andersen[58], 0.5 fs time step). From the trajectories, 500k points were randomly chosen to train the model with a batch size of 64, and 8192 points were also randomly chosen for validation (4 batches tested per epoch). The same initial conditions were also used to test the interatomic distance distribution.

*Virtual particles with L-J potential in PBC*

A total number of 4010-ps NVE trajectories (0.1 fs time step) of the system described in the section *Periodic systems* were generated with initial conditions incubated at 300 K for 1 ps (Nosé–Hoover, 0.5 fs time step). 32 trajectories were randomly chosen for training, while the remaining trajectories were used for the comparison of the radial distributions. The Lennard-Jones potential used for the interactions between particle pairs utilizes the following formula:

$$V_{LJ}(r) = 4\varepsilon[\left(\frac{\sigma}{r}\right)^{12} - \left(\frac{\sigma}{r}\right)^{6}], \quad (5)$$

where $r$ is the distance between two particles. The parameters $\varepsilon$ and $\sigma$ used for each particle pair are provided in Table S2.

From the training trajectories, 500k points were randomly chosen to train the model with a batch size of 160, and 8192 points were also randomly chosen for validation (4 batches tested per epoch).

***Propagation***

The propagation of MD trajectories with the MDtrajNet models can be done with time segments instead of time steps in traditional MD. Any value within the time cutoff used in the training can be used as the length of the time segment. Inside each segment, predictions of positions at any time frame can be made simultaneously, while the predictions of the velocities can be made via either numerical or analytical differentiation. The prediction at the end of the segment can then be used as the initial conditions for the next segment.

In this work, 8 fs was chosen for all the models trained with a 10-fs time cutoff. For models targeting the trajectories in the NVE ensemble, the velocities in the new initial conditions were uniformly rescaled to maintain the total energy, which required an extra evaluation of the reference potential energy surface per segment. For the ethanol's NVT model, the last points of predicted segments were directly used as the new initial conditions, without any rescaling.

**Data availability**

The data is available under the MIT license at https://doi.org/10.5281/zenodo.15588975.

**Code availability**

The code of MDtrajNet and the foundational model MDtrajNet-1 are available in the open-source MLatom at https://github.com/dralgroup/mlatom. Simulations with MDtrajNet can also be performed online on the Aitomistic Hub (https://aitomistic.xyz) and Aitomistic Lab@XMU (https://atom.xmu.edu.cn, free for academic users), packed with AI agents assisting in atomistic simulations.

**Author contributions**

F. G: methodology, software, validation, formal analysis, investigation, data curation, visualization, writing – original draft, writing – review & editing.

P.O.D: conceptualization, methodology, resources, writing – review & editing, supervision, project administration, funding acquisition.

Y. C: revision of MDtrajNet-1 benchmarks, fitting of the reconstructed ANI on the MDtrajNet-1 training dataset and comparative benchmarking of this model, MDtrajNet-1, and sub-models of ANI-1ccx on the MD performance, supplementary video for exemplary trajectory evolution.

**Acknowledgments**

P.O.D. acknowledges funding by the projects for International Senior Scientists (Project No.: W2531013) and for Outstanding Youth Scholars (Overseas, 2021) of the National Natural Science Foundation of China, and via the Lab project of the State Key Laboratory of Physical Chemistry of Solid Surfaces. P.O.D. also thanks Aitomistic, Shenzhen, for funding.

**Supporting Information**

Supporting information: Additional results of MDtrajNet learning on single atomistic systems (Figures S1–S7 and Table S1) and fine-tuning to alanine (Figures S8 and S9). Plot showing root-mean-square error (RMSE) and $R^2$ of different models on 1-ps trajectories of 200 testing system (Figures S10 and S11). Plot for distribution of training data (Figure S12). Plot for comparison between MDtrajNet-1 and ANI-1x on 10-ps long trajectories (Figure S13). Table for parameters of Lennard-Jones potential used in this work (Table S2).

Supporting information: Movie for example trajectories propagated with MDtrajNet-1, reference ANI-1ccx, and reconstructed MLIP.

**References**

(1) Griffiths, D. J.; Schroeter, D. F., *Introduction to Quantum Mechanics*. 3 ed.; Cambridge University Press: Cambridge, 2018.

(2) Rapaport, D. C., *The Art of Molecular Dynamics Simulation*. 2 ed.; Cambridge University Press: Cambridge, 2004.

(3) Alder, B. J.; Wainwright, T. E., Studies in Molecular Dynamics. I. General Method. *The Journal of Chemical Physics* **1959,** *31*, 459-466.

(4) Tuckerman, M. E.; Martyna, G. J., Understanding Modern Molecular Dynamics: Techniques and Applications. *The Journal of Physical Chemistry B* **2000,** *104*, 159-178.

(5) Scheraga, H. A.; Khalili, M.; Liwo, A., Protein-Folding Dynamics: Overview of Molecular Simulation Techniques. *Annual Review of Physical Chemistry* **2007,** *58*, 57-83.

(6) Gaigeot, M.-P.; Michaël, M.; and Vuilleumier, R., Infrared spectroscopy in the gas and liquid phase from first principle molecular dynamics simulations: application to small peptides. *Molecular Physics* **2007,** *105*, 2857-2878.

(7) Gissinger, J. R.; Jensen, B. D.; Wise, K. E., Modeling chemical reactions in classical molecular dynamics simulations. *Polymer* **2017,** *128*, 211-217.

(8) Sun, Y.; Kollman, P. A., Conformational sampling and ensemble generation by molecular dynamics simulations: 18-Crown-6 as a test case. *Journal of Computational Chemistry* **1992,** *13*, 33-40.

(9) Bruccoleri, R. E.; Karplus, M., Conformational sampling using high-temperature molecular dynamics. *Biopolymers* **1990,** *29*, 1847-1862.

(10) Chmiela, S.; Sauceda, H. E.; Müller, K.-R.; Tkatchenko, A., Towards exact molecular dynamics simulations with machine-learned force fields. *Nature Communications* **2018,** *9*, 3887.

(11) Zhong, X.; Zhao, Y., Chapter 5 - Basics of dynamics. In *Quantum Chemistry in the Age of Machine Learning*, Dral, P. O., Ed. Elsevier: 2023; pp 117-133.

(12) Dral, P. O., Quantum Chemistry in the Age of Machine Learning. *The Journal of Physical Chemistry Letters* **2020,** *11*, 2336-2347.

(13) Unke, O. T.; Chmiela, S.; Sauceda, H. E.; Gastegger, M.; Poltavsky, I.; Schütt, K. T.; Tkatchenko, A.; Müller, K.-R., Machine Learning Force Fields. *Chemical Reviews* **2021,** *121*, 10142-10186.

(14) Behler, J., Four Generations of High-Dimensional Neural Network Potentials. *Chemical Reviews* **2021,** *121*, 10037-10072.

(15) Ge, F.; Zhang, L.; Hou, Y.-F.; Chen, Y.; Ullah, A.; Dral, P. O., Four-Dimensional-Spacetime Atomistic Artificial Intelligence Models. *The Journal of Physical Chemistry Letters* **2023,** *14*, 7732-7743.

(16) Wang, J.; Li, C.; Shin, S.; Qi, H., Accelerated Atomic Data Production in Ab Initio Molecular Dynamics with Recurrent Neural Network for Materials Research. *J. Phys. Chem. C* **2020,** *124*, 14838–14846.

(17) Zheng, T.; Gao, W.; Wang, C., Learning large-time-step molecular dynamics with graph neural networks. *arXiv preprint arXiv:2111.15176* **2021**.

(18) Bigi, F.; Chong, S.; Kristiadi, A.; Ceriotti, M., FlashMD: long-stride, universal prediction of molecular dynamics. *arXiv preprint arXiv:2505.19350* **2025**.

(19) Diez, J. V.; Schreiner, M.; Engkvist, O.; Olsson, S., Boltzmann priors for Implicit Transfer Operators. *arXiv:2410.10605v2 [physics.chem-ph]* **2025**.

(20) Vaswani, A.; Shazeer, N.; Parmar, N.; Uszkoreit, J.; Jones, L.; Gomez, A. N.; Kaiser, Ł.; Polosukhin, I., Attention is all you need. In *Proceedings of the 31st International Conference on Neural Information Processing Systems*, Curran Associates Inc.: Long Beach, California, USA, 2017; pp 6000–6010.

(21) OpenAi; Achiam, J.; Adler, S.; Agarwal, S.; Ahmad, L.; Akkaya, I.; Aleman, F. L.; Almeida, D.; Altenschmidt, J.; Altman, S.; Anadkat, S.; Avila, R.; Babuschkin, I.; Balaji, S.; Balcom, V.; Baltescu, P.; Bao, H.; Bavarian, M.; Belgum, J.; Bello, I.; Berdine, J.; Bernadett-Shapiro, G.; Berner, C.; Bogdonoff, L.; Boiko, O.; Boyd, M.; Brakman, A.-L.; Brockman, G.; Brooks, T.; Brundage, M.; Button, K.; Cai, T.; Campbell, R.; Cann, A.; Carey, B.; Carlson, C.; Carmichael, R.; Chan, B.; Chang, C.; Chantzis, F.; Chen, D.; Chen, S.; Chen, R.; Chen, J.; Chen, M.; Chess, B.; Cho, C.; Chu, C.; Chung, H. W.; Cummings, D.; Currier, J.; Dai, Y.; Decareaux, C.; Degry, T.; Deutsch, N.; Deville, D.; Dhar, A.; Dohan, D.; Dowling, S.; Dunning, S.; Ecoffet, A.; Eleti, A.; Eloundou, T.; Farhi, D.; Fedus, L.; Felix, N.; Fishman, S. P.; Forte, J.; Fulford, I.; Gao, L.; Georges, E.; Gibson, C.; Goel, V.; Gogineni, T.; Goh, G.; Gontijo-Lopes, R.; Gordon, J.; Grafstein, M.; Gray, S.; Greene,

R.; Gross, J.; Gu, S. S.; Guo, Y.; Hallacy, C.; Han, J.; Harris, J.; He, Y.; Heaton, M.; Heidecke, J.; Hesse, C.; Hickey, A.; Hickey, W.; Hoeschele, P.; Houghton, B.; Hsu, K.; Hu, S.; Hu, X.; Huizinga, J.; Jain, S.; Jain, S.; Jang, J.; Jiang, A.; Jiang, R.; Jin, H.; Jin, D.; Jomoto, S.; Jonn, B.; Jun, H.; Kaftan, T.; Kaiser, Ł.; Kamali, A.; Kanitscheider, I.; Keskar, N. S.; Khan, T.; Kilpatrick, L.; Kim, J. W.; Kim, C.; Kim, Y.; Kirchner, J. H.; Kiros, J.; Knight, M.; Kokotajlo, D.; Kondraciuk, Ł.; Kondrich, A.; Konstantinidis, A.; Kosic, K.; Krueger, G.; Kuo, V.; Lampe, M.; Lan, I.; Lee, T.; Leike, J.; Leung, J.; Levy, D.; Li, C. M.; Lim, R.; Lin, M.; Lin, S.; Litwin, M.; Lopez, T.; Lowe, R.; Lue, P.; Makanju, A.; Malfacini, K.; Manning, S.; Markov, T.; Markovski, Y.; Martin, B.; Mayer, K.; Mayne, A.; McGrew, B.; McKinney, S. M.; McLeavey, C.; McMillan, P.; McNeil, J.; Medina, D.; Mehta, A.; Menick, J.; Metz, L.; Mishchenko, A.; Mishkin, P.; Monaco, V.; Morikawa, E.; Mossing, D.; Mu, T.; Murati, M.; Murk, O.; Mély, D.; Nair, A.; Nakano, R.; Nayak, R.; Neelakantan, A.; Ngo, R.; Noh, H.; Ouyang, L.; O'Keefe, C.; Pachocki, J.; Paino, A.; Palermo, J.; Pantuliano, A.; Parascandolo, G.; Parish, J.; Parparita, E.; Passos, A.; Pavlov, M.; Peng, A.; Perelman, A.; Peres, F. d. A. B.; Petrov, M.; Pinto, H. P. d. O.; Michael; Pokorny; Pokrass, M.; Pong, V. H.; Powell, T.; Power, A.; Power, B.; Proehl, E.; Puri, R.; Radford, A.; Rae, J.; Ramesh, A.; Raymond, C.; Real, F.; Rimbach, K.; Ross, C.; Rotsted, B.; Roussez, H.; Ryder, N.; Saltarelli, M.; Sanders, T.; Santurkar, S.; Sastry, G.; Schmidt, H.; Schnurr, D.; Schulman, J.; Selsam, D.; Sheppard, K.; Sherbakov, T.; Shieh, J.; Shoker, S.; Shyam, P.; Sidor, S.; Sigler, E.; Simens, M.; Sitkin, J.; Slama, K.; Sohl, I.; Sokolowsky, B.; Song, Y.; Staudacher, N.; Such, F. P.; Summers, N.; Sutskever, I.; Tang, J.; Tezak, N.; Thompson, M. B.; Tillet, P.; Tootoonchian, A.; Tseng, E.; Tuggle, P.; Turley, N.; Tworek, J.; Uribe, J. F. C.; Vallone, A.; Vijayvergiya, A.; Voss, C.; Wainwright, C.; Wang, J. J.; Wang, A.; Wang, B.; Ward, J.; Wei, J.; Weinmann, C. J.; Welihinda, A.; Welinder, P.; Weng, J.; Weng, L.; Wiethoff, M.; Willner, D.; Winter, C.; Wolrich, S.; Wong, H.; Workman, L.; Wu, S.; Wu, J.; Wu, M.; Xiao, K.; Xu, T.; Yoo, S.; Yu, K.; Yuan, Q.; Zaremba, W.; Zellers, R.; Zhang, C.; Zhang, M.; Zhao, S.; Zheng, T.; Zhuang, J.; Zhuk, W.; Zoph, B., GPT-4 Technical Report. *arXiv [cs.CL]* **2023**.

(22) Devlin, J.; Chang, M.-W.; Lee, K.; Toutanova, K., BERT: Pre-training of Deep Bidirectional Transformers for Language Understanding. 2019; pp 4171-4186.

(23) El-Nouby, A.; Touvron, H.; Caron, M.; Bojanowski, P.; Douze, M.; Joulin, A.; Laptev, I.; Neverova, N.; Synnaeve, G.; Verbeek, J.; Jégou, H., XCiT: cross-covariance image transformers. In *Proceedings of the 35th International Conference on Neural Information Processing Systems*, Curran Associates Inc.: 2021; p Article 1531.

(24) Jamil, S.; Piran, M. J.; Kwon, O.-J., A comprehensive survey of transformers for computer vision. *arXiv [cs.CV]* **2022**.

(25) Heo, B.; Yun, S.; Han, D.; Chun, S.; Choe, J.; Oh, S. J. In *Rethinking Spatial Dimensions of Vision Transformers*, 2021 IEEE/CVF International Conference on Computer Vision (ICCV), 10-17 Oct. 2021; 2021; pp 11916-11925.

(26) Zhou, G.; Gao, Z.; Ding, Q.; Zheng, H.; Xu, H.; Wei, Z.; Zhang, L.; Ke, G. In *Uni-mol: A universal 3d molecular representation learning framework*, The Eleventh International Conference on Learning Representations, 2023.

(27) Lu, S.; Gao, Z.; He, D.; Zhang, L.; Ke, G., Data-driven quantum chemical property prediction leveraging 3D conformations with Uni-Mol+. *Nature Communications* **2024,** *15*, 7104.

(28) Janson, G.; Valdes-Garcia, G.; Heo, L.; Feig, M., Direct generation of protein conformational ensembles via machine learning. *Nature Communications* **2023,** *14*, 774.

(29) Ilyes, B.; David Peter, K.; Gregor, N. C. S.; Christoph, O.; Gabor, C., MACE : Higher Order Equivariant Message Passing Neural Networks for Fast and Accurate Force Fields. Alice, H. O.; Alekh, A.; Danielle, B.; Kyunghyun, C., Eds. 2022.

(30) Batzner, S.; Musaelian, A.; Sun, L.; Geiger, M.; Mailoa, J. P.; Kornbluth, M.; Molinari, N.; Smidt, T. E.; Kozinsky, B., E(3)-equivariant graph neural networks for data-efficient and accurate interatomic potentials. *Nature Communications* **2022,** *13*, 2453.

(31) Musaelian, A.; Batzner, S.; Johansson, A.; Sun, L.; Owen, C. J.; Kornbluth, M.; Kozinsky, B., Learning local equivariant representations for large-scale atomistic dynamics. *Nature Communications* **2023,** *14*, 579.

(32) Liu, Z.; Yang, D.; Wang, Y.; Lu, M.; Li, R., EGNN: Graph structure learning based on evolutionary computation helps more in graph neural networks. *Appl. Soft Comput.* **2023,** *135*, 12.

(33) Schütt, K. T.; Unke, O. T.; Gastegger, M., Equivariant message passing for the prediction of tensorial properties and molecular spectra. *arXiv [cs.LG]* **2021**.

(34) Haghighatlari, M.; Li, J.; Guan, X.; Zhang, O.; Das, A.; Stein, C. J.; Heidar-Zadeh, F.; Liu, M.; Head-Gordon, M.; Bertels, L.; Hao, H.; Leven, I.; Head-Gordon, T., NewtonNet: a Newtonian message passing network for deep learning of interatomic potentials and forces. *Digital Discovery* **2022,** *1*, 333-343.

(35) Lin, C. E.; Zhu, M.; Ghaffari, M., SE3ET: SE(3)-Equivariant Transformer for Low-Overlap Point Cloud Registration. *IEEE Robotics and Automation Letters* **2024,** *9*, 9526-9533.

(36) Tai, K. S.; Bailis, P.; Valiant, G., Equivariant Transformer Networks. *arXiv [cs.CV]* **2019**.

(37) Philipp, T.; Gianni De, F., Equivariant Transformers for Neural Network based Molecular Potentials. In *International Conference on Learning Representations*, 2022.

(38) Pelaez, R. P.; Simeon, G.; Galvelis, R.; Mirarchi, A.; Eastman, P.; Doerr, S.; Thölke, P.; Markland, T. E.; De Fabritiis, G., TorchMD-Net 2.0: Fast Neural Network Potentials for Molecular Simulations. *Journal of Chemical Theory and Computation* **2024,** *20*, 4076-4087.

(39) Yang, Z.; Zeng, X.; Zhao, Y.; Chen, R., AlphaFold2 and its applications in the fields of biology and medicine. *Signal Transduction and Targeted Therapy* **2023,** *8*, 115.

(40) Fuchs, F. B.; Worrall, D. E.; Fischer, V.; Welling, M., SE(3)-transformers: 3D roto-translation equivariant attention networks. In *Proceedings of the 34th International Conference on Neural Information Processing Systems*, Curran Associates Inc.: Vancouver, BC, Canada, 2020; p Article 166.

(41) Smith, J. S.; Nebgen, B.; Lubbers, N.; Isayev, O.; Roitberg, A. E., Less is more: Sampling chemical space with active learning. *The Journal of Chemical Physics* **2018,** *148*.

(42) Janoš, J.; Slavíček, P., What Controls the Quality of Photodynamical Simulations? Electronic Structure Versus Nonadiabatic Algorithm. *Journal of Chemical Theory and Computation* **2023,** *19*, 8273-8284.

(43) Young, T. A.; Johnston-Wood, T.; Zhang, H.; Duarte, F., Reaction dynamics of Diels–Alder reactions from machine learned potentials. *Physical Chemistry Chemical Physics* **2022,** *24*, 20820-20827.

(44) Zhang, L.; Hou, Y.-F.; Ge, F.; Dral, P. O., Energy-conserving molecular dynamics is not energy conserving. *Physical Chemistry Chemical Physics* **2023,** *25*, 23467-23476.

(45) Zhang, L.; Hou, Y.-F.; Ge, F.; Dral, P. O., Energy-conserving molecular dynamics is not energy conserving. *Phys. Chem. Chem. Phys.* **2023,** *25*, 23467–23476.

(46) Esch, B. V.; Peters, L. D. M.; Sauerland, L.; Ochsenfeld, C., Quantitative Comparison of Experimental and Computed IR-Spectra Extracted from Ab Initio Molecular Dynamics. *J. Chem. Theory Comput.* **2021,** *17*, 985–995.

(47) Nyquist, H., Certain Topics in Telegraph Transmission Theory. *Transactions of the American Institute of Electrical Engineers* **1928,** *47*, 617-644.

(48) Shannon, C. E., Communication in the Presence of Noise. *Proceedings of the IRE* **1949,** *37*, 10-21.

(49) Mironov, V.; Alexeev, Y.; Mulligan, V. K.; Fedorov, D. G., A systematic study of minima in alanine dipeptide. *Journal of Computational Chemistry* **2019,** *40*, 297-309.

(50) Zhang, S.; Makoś, M. Z.; Jadrich, R. B.; Kraka, E.; Barros, K.; Nebgen, B. T.; Tretiak, S.; Isayev, O.; Lubbers, N.; Messerly, R. A.; Smith, J. S., Exploring the frontiers of condensed-phase chemistry with a general reactive machine learning potential. *Nature Chemistry* **2024,** *16*, 727-734.

(51) Geiger, M.; Smidt, T., e3nn: Euclidean Neural Networks. *arXiv [cs.LG]* **2022**.

(52) Dral, P. O.; Ge, F.; Hou, Y.-F.; Zheng, P.; Chen, Y.; Barbatti, M.; Isayev, O.; Wang, C.; Xue, B.-X.; Pinheiro Jr, M.; Su, Y.; Dai, Y.; Chen, Y.; Zhang, L.; Zhang, S.; Ullah, A.; Zhang, Q.; Ou, Y., MLatom 3: A Platform for Machine Learning-Enhanced Computational Chemistry Simulations and Workflows. *Journal of Chemical Theory and Computation* **2024,** *20*, 1193-1213.

(53) Dahl, J. P.; Springborg, M., The Morse oscillator in position space, momentum space, and phase space. *The Journal of Chemical Physics* **1988,** *88*, 4535-4547.

(54) Schinke, R., *Photodissociation Dynamics: Spectroscopy and Fragmentation of Small Polyatomic Molecules*. Cambridge University Press: Cambridge, 1993.

(55) Nosé, S., A unified formulation of the constant temperature molecular dynamics methods. *The Journal of Chemical Physics* **1984,** *81*, 511-519.

(56) Hoover, W. G., Canonical dynamics: Equilibrium phase-space distributions. *Physical Review A* **1985,** *31*, 1695-1697.

(57) O'Boyle, N. M.; Banck, M.; James, C. A.; Morley, C.; Vandermeersch, T.; Hutchison, G. R., Open Babel: An open chemical toolbox. *Journal of Cheminformatics* **2011,** *3*, 33.

(58) Andersen, H. C., Molecular dynamics simulations at constant pressure and/or temperature. *The Journal of Chemical Physics* **1980,** *72*, 2384-2393.

**TOC Graphic**

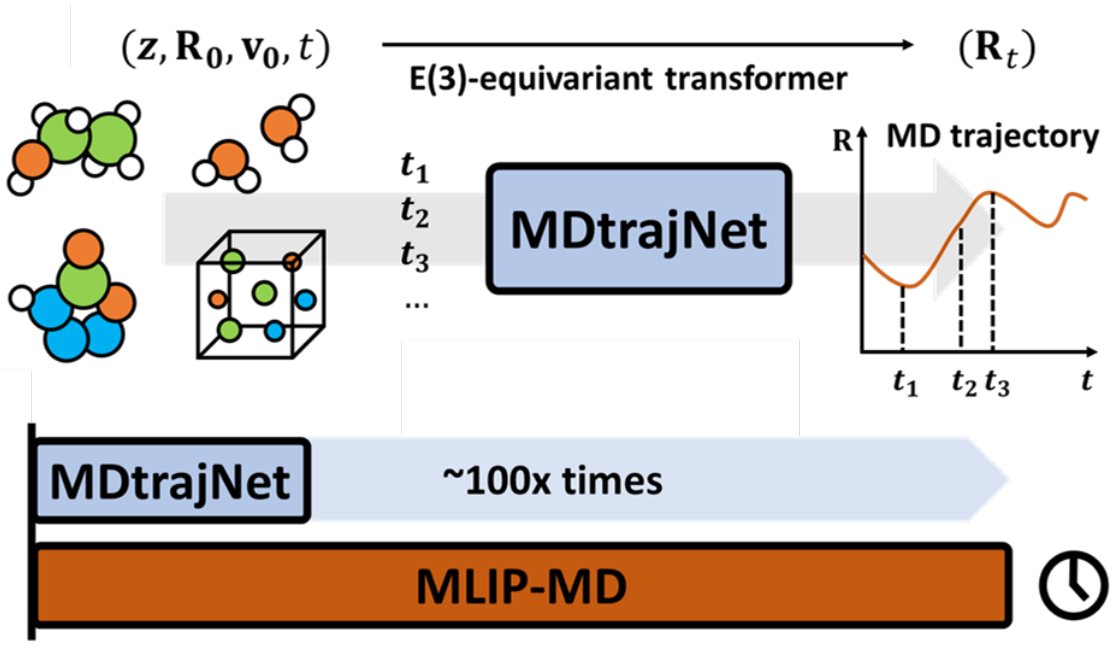

$(z, \mathbf{R}_0, \mathbf{v}_0, t)$
E(3)-equivariant transformer
$(\mathbf{R}_t)$
$t_1$
$t_2$
$t_3$
...
MDtrajNet
R
MD trajectory
$t_1$ $t_2$ $t_3$ $t$
MDtrajNet
~100x times
MLIP-MD

# Supporting Information for
# Artificial Intelligence for Direct Prediction of Molecular Dynamics Across Chemical Space

Fuchun Ge,[1] Yuxinxin Chen,[1,3] and Pavlo O. Dral[1,2,3*]

[1]*State Key Laboratory of Physical Chemistry of Solid Surfaces, Department of Chemistry, College of Chemistry and Chemical Engineering, and Fujian Provincial Key Laboratory of Theoretical and Computational Chemistry, Xiamen University, Xiamen 361005, China.*

[2]*Institute of Physics, Faculty of Physics, Astronomy, and Informatics, Nicolaus Copernicus University in Toruń, ul. Grudziądzka 5, 87-100 Toruń, Poland.*

[3]*Aitomistic, Shenzhen 518000, China.*

*Email: dral@xmu.edu.cn*

**Table of Contents**

***Performance of the MDtrajNet on learning a single atomistic system***

To evaluate the MDtrajNet as a new implementation of the 4D-spacetime model for dynamic molecular systems, we compared its performance against the original GICnet implementation[1] on a single atomistic system, for which the GICnet model is limited. The monomolecular system of ethanol, which was extensively discussed in the previous work introducing GICnet, was examined to highlight the differences between the two approaches.

The prediction error within a time segment serves as the primary performance metric and is directly used during model training. The MDtrajNet model demonstrates an exceptional ability to generalize atomic motion into the near future, achieving sub-picometer accuracy. For instance, as shown in Figure S1, the model attained a test RMSE of 0.22 pm for the Cartesian coordinates of atoms in the ethanol molecule over a 10-fs interval—an error on the order of a thousandth of typical chemical bond lengths in such a molecule. Compared to GICnet, which was trained on the same dataset, MDtrajNet exhibits significantly improved generalization, nearly doubling the accuracy of position and velocity predictions. This indicates the improved architecture design of MDtrajNet.

Within time segments, average prediction errors vary across atoms and time points (Figure S3). Larger errors tend to occur at longer time intervals and for lighter atoms, consistent with the intuition that more flexible atomic motions are harder to model. Another notable contributor to prediction error is the "average sparsity" of training data in the input space, particularly at the edges of the 0–10 fs window, where fewer neighboring data points exist. While this effect is subtle in geometry predictions, it becomes more pronounced in their time derivatives, leading to basin-shaped RMSE distributions for velocities. This highlights the importance of having adjacent time frames in learning classical dynamics, whose deterministic nature leads to a high correlation from frame to frame. This insight also explains why training on the full segment yields better performance than training solely on a part of it (Figure S4).

**Table S1.** The mean test RMSEs within 10 fs of two MDtrajNet models trained on ethanol NVE and NVT trajectories, respectively. For the Cartesian system, values are calculated by averaging the root-mean-square error on each Cartesian component. For bond-angle-torsion, the RMSE values are calculated only for the coordinates present in the Z-matrix before averaging. Values in parentheses are the mean test RMSEs from the GICnet model of ethanol.

| | | Cartesian (pm) | bond length (pm) | bond angle (centiradian) | torsion angle (centiradian) |
|---|---|---|---|---|---|
| NVE | value | 0.220 | 0.229 (0.394) | 0.326 (0.882) | 0.368 (0.984) |
| | time derivative ($fs^{-1}$) | 0.111 | 0.145 (0.178) | 0.151 (0.220) | 0.170 (0.244) |
| NVT | value | 0.341 | 0.237 | 0.499 | 0.669 |
| | time derivative ($fs^{-1}$) | 0.115 | 0.140 | 0.156 | 0.196 |

centiradian: 100th of a radian.

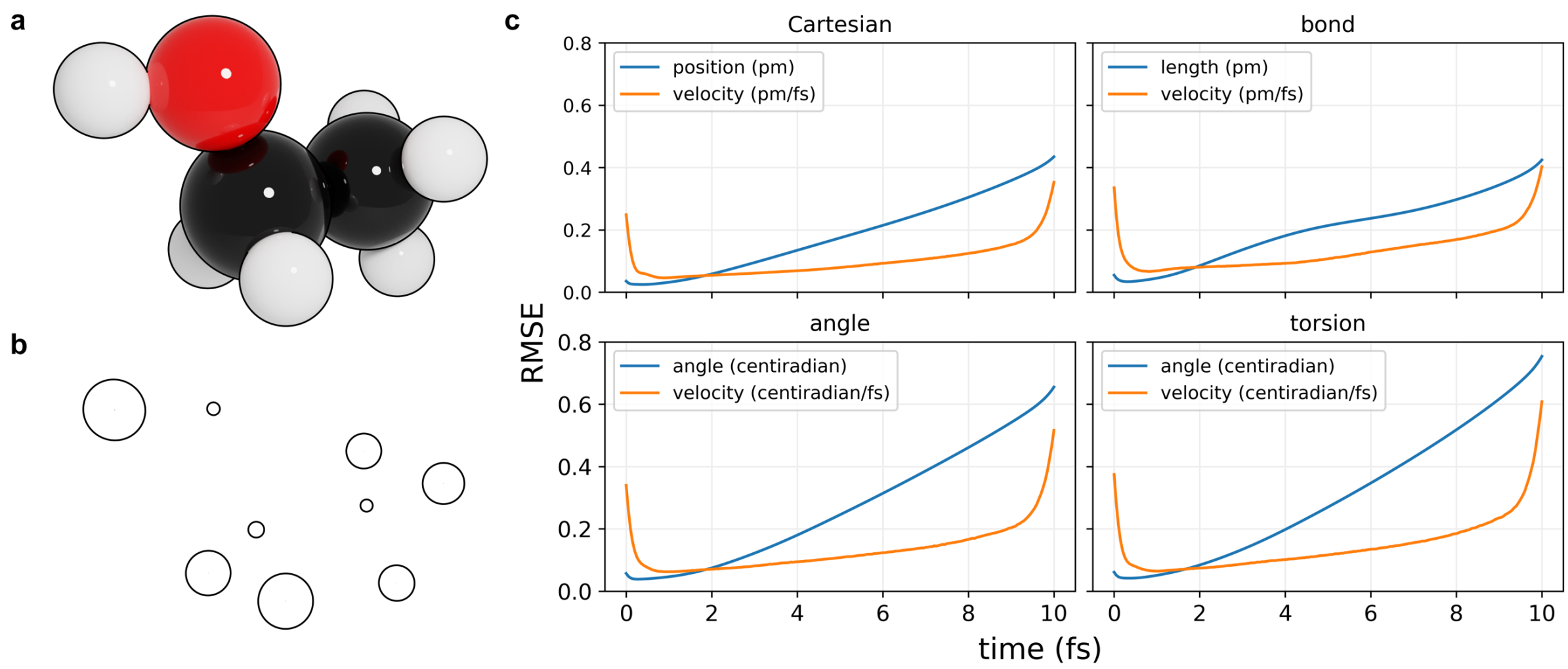

**Figure S1.** a. A 3D graphic representation of the ethanol molecule, with atoms labeled with identifiers. b. Visualized in-segment test errors of the atom positions of the ethanol monomolecular system. The barely visible spheres are plotted with their radii set to the root-mean-square norm of the error in corresponding atom positions, while the circular outlines show those errors scaled 100 times up. 3D objects are shown with the same viewing perspective as in part a. c. In-segment model test error versus time for ethanol in Cartesian and internal coordinate systems. The errors of the predicted position are shown in blue, with a picometer set as the unit for Cartesian components and the bond lengths, and a centiradian (hundredth of a radian) for both bond and torsion angles. The errors of velocities (position differences per fs) are plotted in orange.

Although the error level is visually negligible within segments (Figure S1b), they can accumulate over iterations, thus potentially destabilizing long-term trajectories. Two approaches can mitigate this issue. The first is to train on larger and denser datasets that offer improved accuracy and broader phase space coverage, reducing the chance of encountering poorly sampled extrapolation regions. The second is to actively correct trajectories during propagation. In our previous work with GICnet, we have employed an ensemble-based strategy that utilizes multiple machine-learning models, where only the predictions that give the minimum total energy shifts are used as new initial conditions.

For the new MDtrajNet model, we found that long-term stability could be achieved by simply using velocity rescaling even with a single model. This is supported by the energy distributions and power spectra from long-time propagation as shown Figure S2. The model reproduces potential and kinetic energy distributions with excellent agreement to the reference method, thereby preserving the energy conservation characteristic of the NVE ensemble used in the reference method. Across various test trajectories, the RMS deviation in total energy ranges from approximately 0.4 to 1.8 kcal/mol, correlating positively with the total energy (Figure S7).

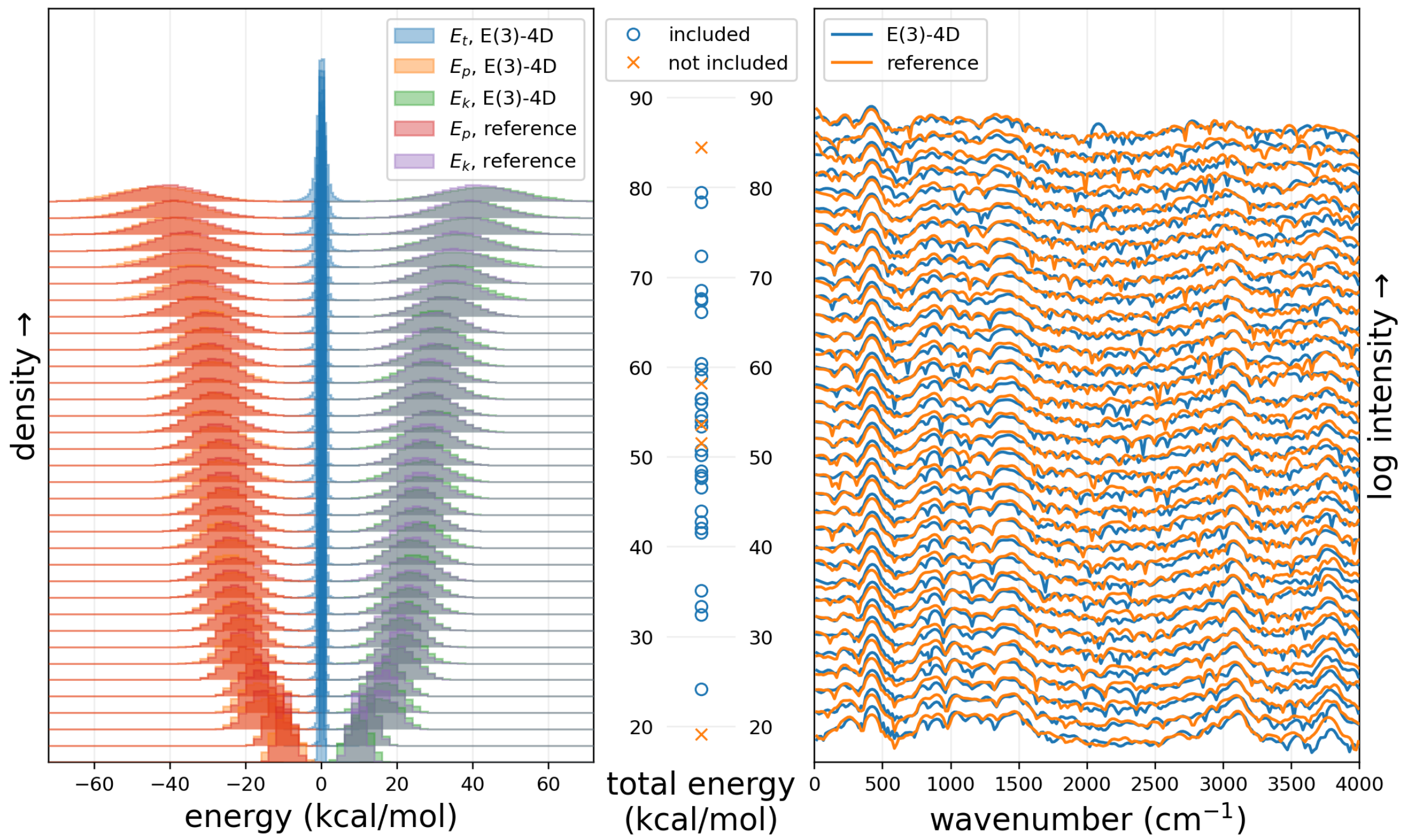

**Figure S2.** Energy distributions (left) and power spectra (right) from a single MDtrajNet model and the reference method for different total energies (middle). Plots that correspond to higher total energies are shifted more towards the top. Total energies in the middle are relative values to the global minimum and those are not included in the training data are marked with crosses. Each histogram is collected from a 10-ps trajectory. Total energies ($E_t$) and potential energies ($E_p$) are relative to the initial total energy while kinetic energies ($E_k$) are not shifted. Each power spectrum is calculated from 32 different 10-ps trajectories with the same total energy. Separated plots for every total energy can be found in Figure S5 and Figure S6.

In terms of vibrational dynamics, the power spectra extracted from long-term trajectories also align closely with the reference method, for both seen and unseen energy regimes. MDtrajNet accurately reproduces nearly all spectral peaks with reliable intensities. Some systematic deviations of intensities are observed—for example, an overestimation for the peak near ~450 $cm^{-1}$. Nevertheless, the model effectively captures the general trend of decreasing spectral intensity above 2500 $cm^{-1}$ as the total energy increases.

These results demonstrate that MDtrajNet not only captures short-term dynamics with high accuracy but also faithfully preserves key energetic and spectral signatures over extended simulations, reinforcing its robustness for long-time molecular dynamics modeling. Notably, the model performs well not only on trajectories within total energies that were seen during training but also on those with previously unseen or even out-of-range ones. This suggests that the model captures the fundamental laws governing atomic motion rather than merely memorizing patterns from training data.

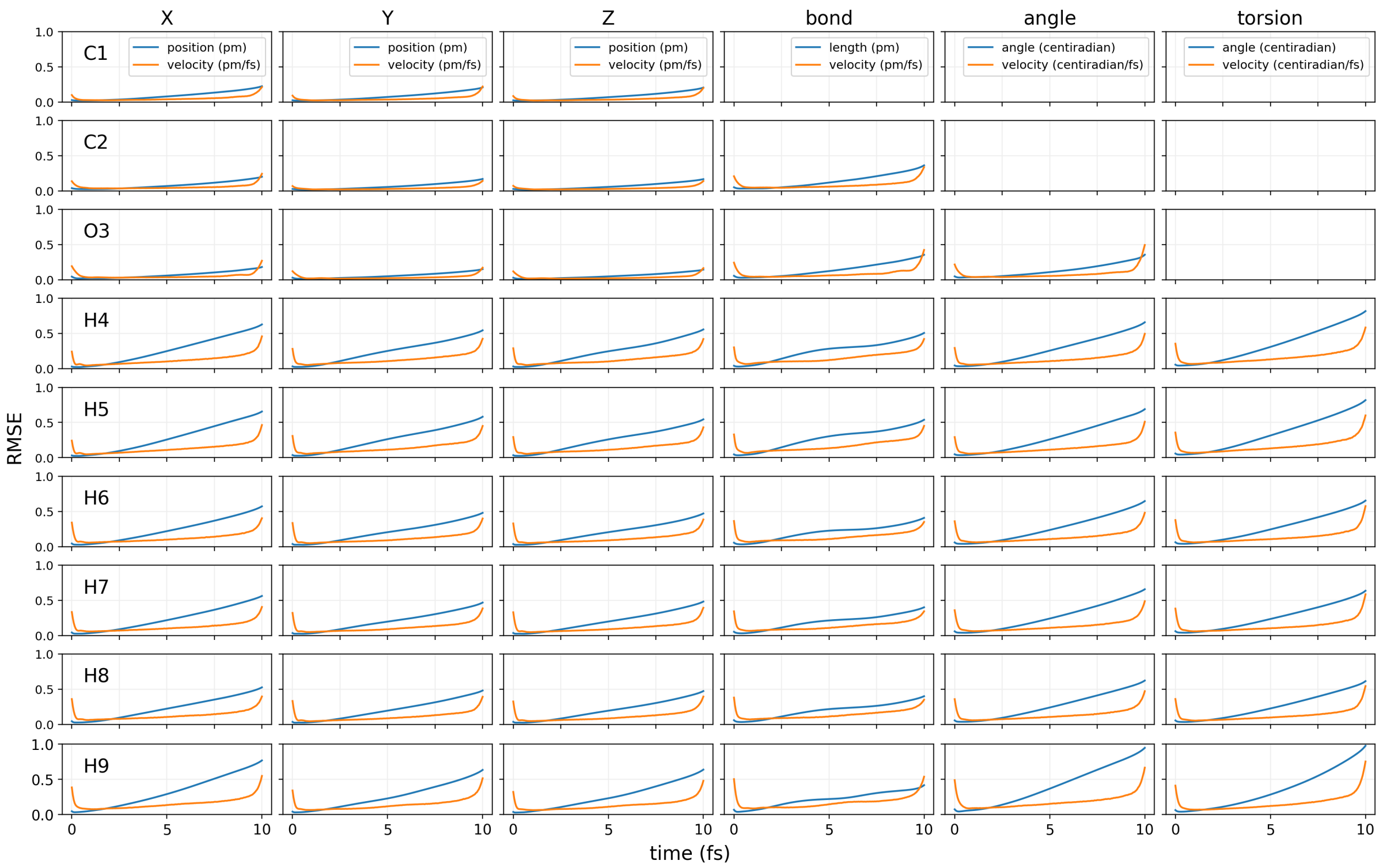

**Figure S3.** In-segment model test error versus time for ethanol in Cartesian and internal coordinates systems. The errors of the predicted position are shown in blue, with a picometer set as the unit for Cartesian components and the bond lengths, and a centiradian (hundredth of a radian) for both bond and torsion angles. The errors of velocities (position differences per fs) are plotted in orange.

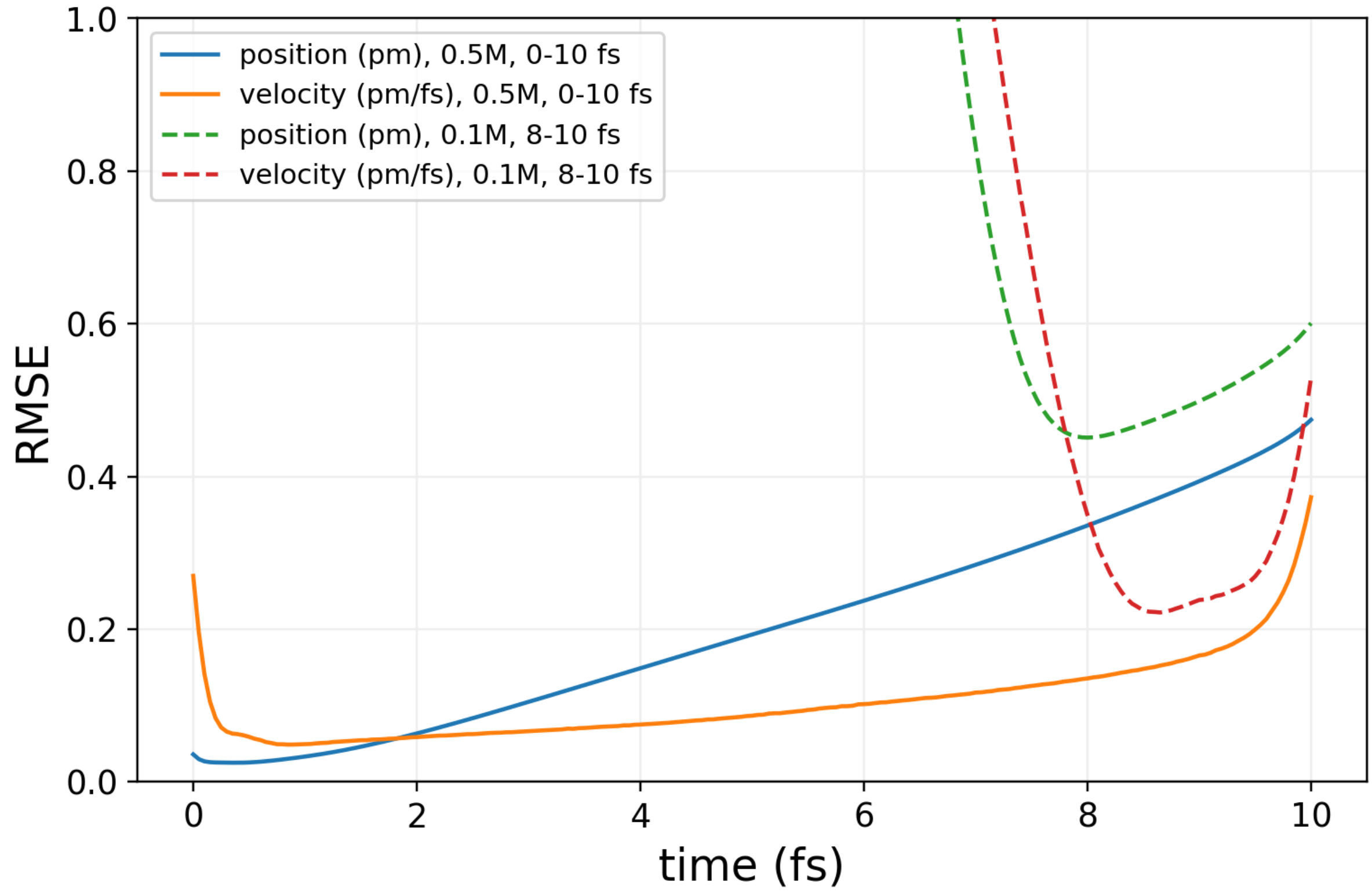

**Figure S4.** RMSEs at different times. The solid lines show the result from the model trained with a dataset of 0.5M points containing data from 0 to 10 fs. The dashed lines show the result from the model trained with a dataset of 0.1M points containing data from 8 to 10 fs. Different times within the specified range are evenly sampled for either dataset.

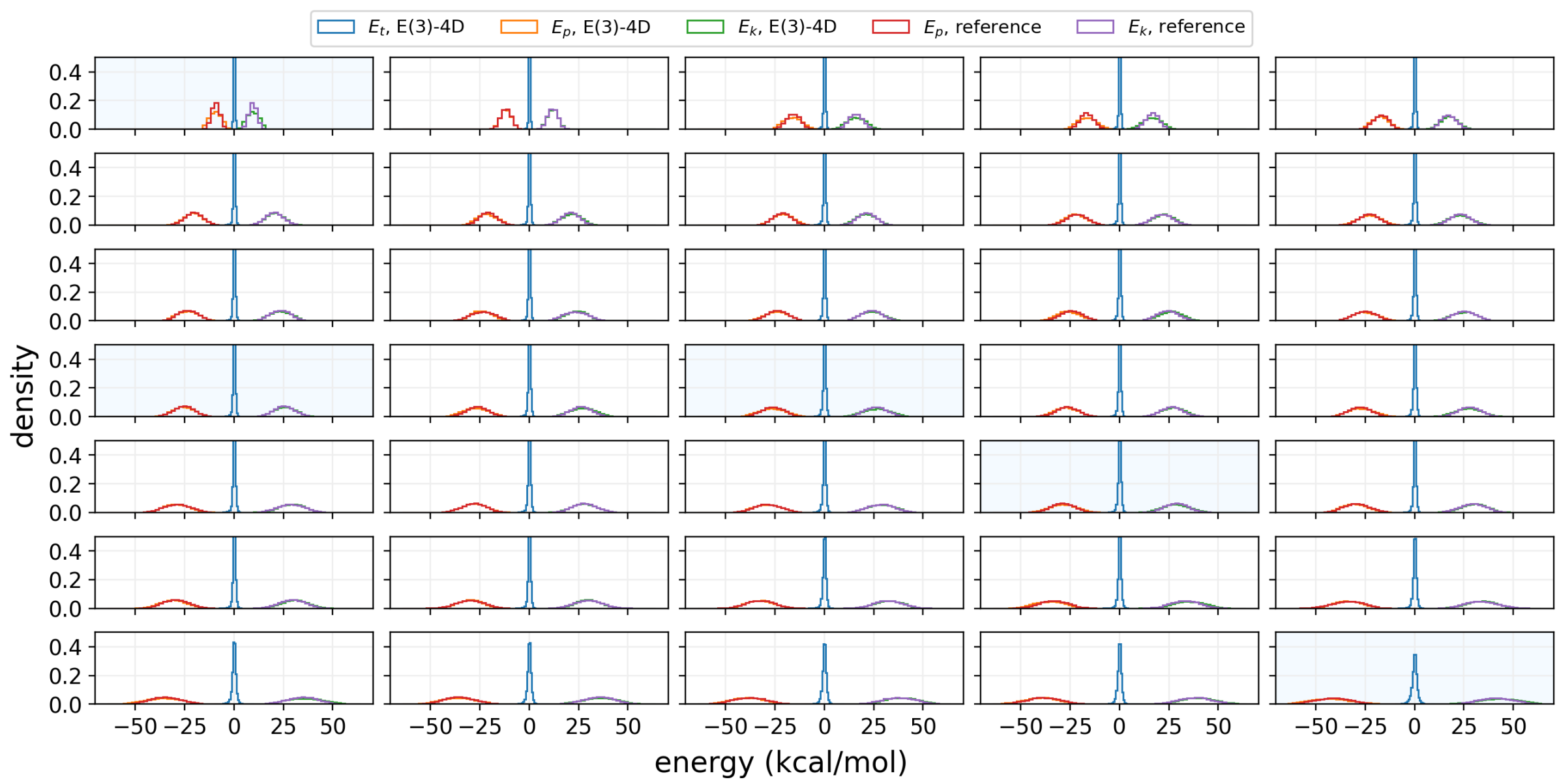

**Figure S5.** Distributions of total energies ($E_t$), potential energies ($E_p$), and kinetic energies ($E_k$) from 10-ps MDtrajNet trajectories compared with references. Initial total energies were subtracted from potential energies and total energies. Histograms are sorted by total energy in ascending order, among those whose corresponding trajectories were excluded from the training set are shown with a light blue background.

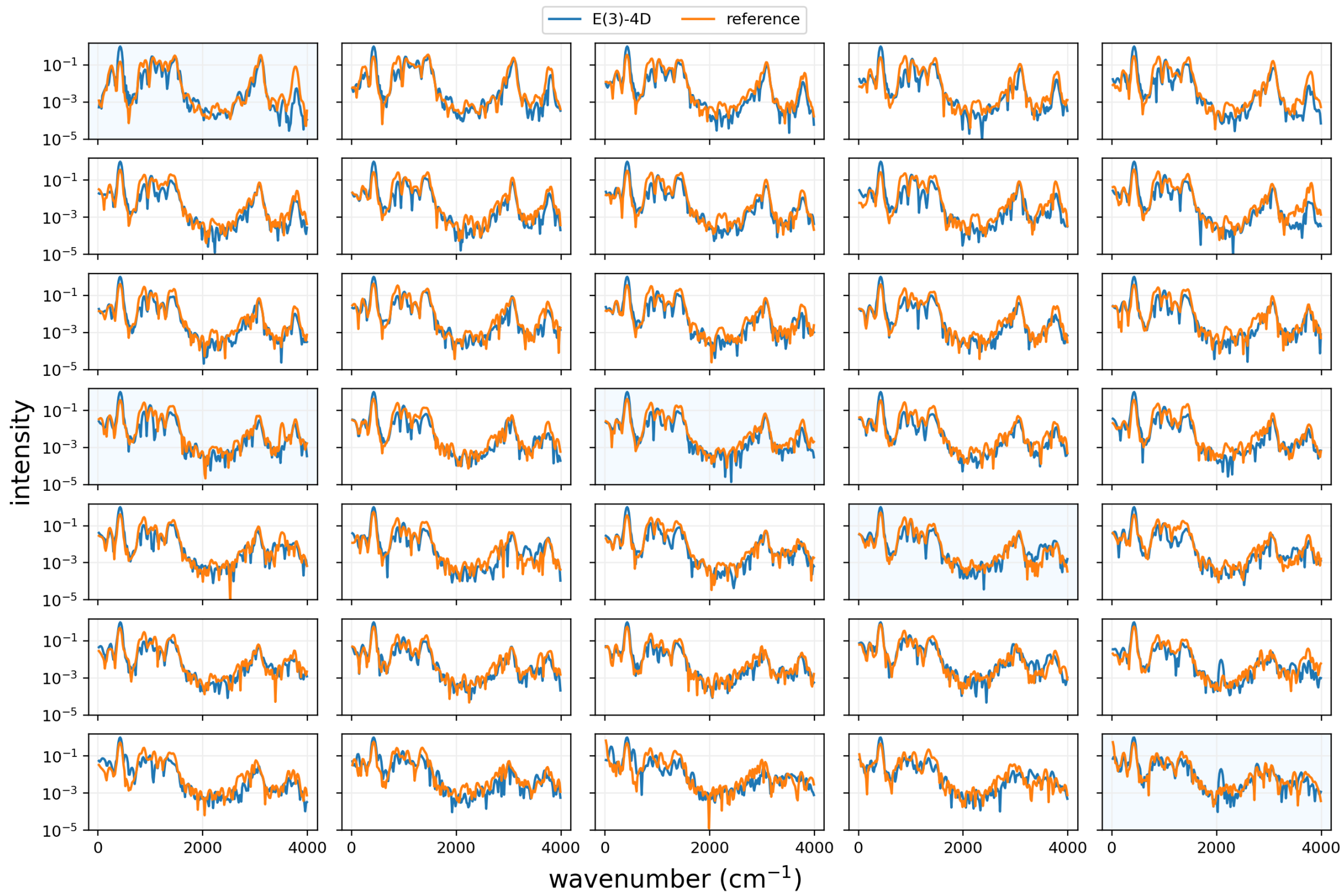

**Figure S6.** Power spectra of the ethanol molecule with different total energies. Each spectrum is generated from 32 different 10-ps trajectories with the same total energy. Spectra are sorted by total energy in ascending order; among those whose corresponding total energies were not included in the training set are shown with a light blue background.

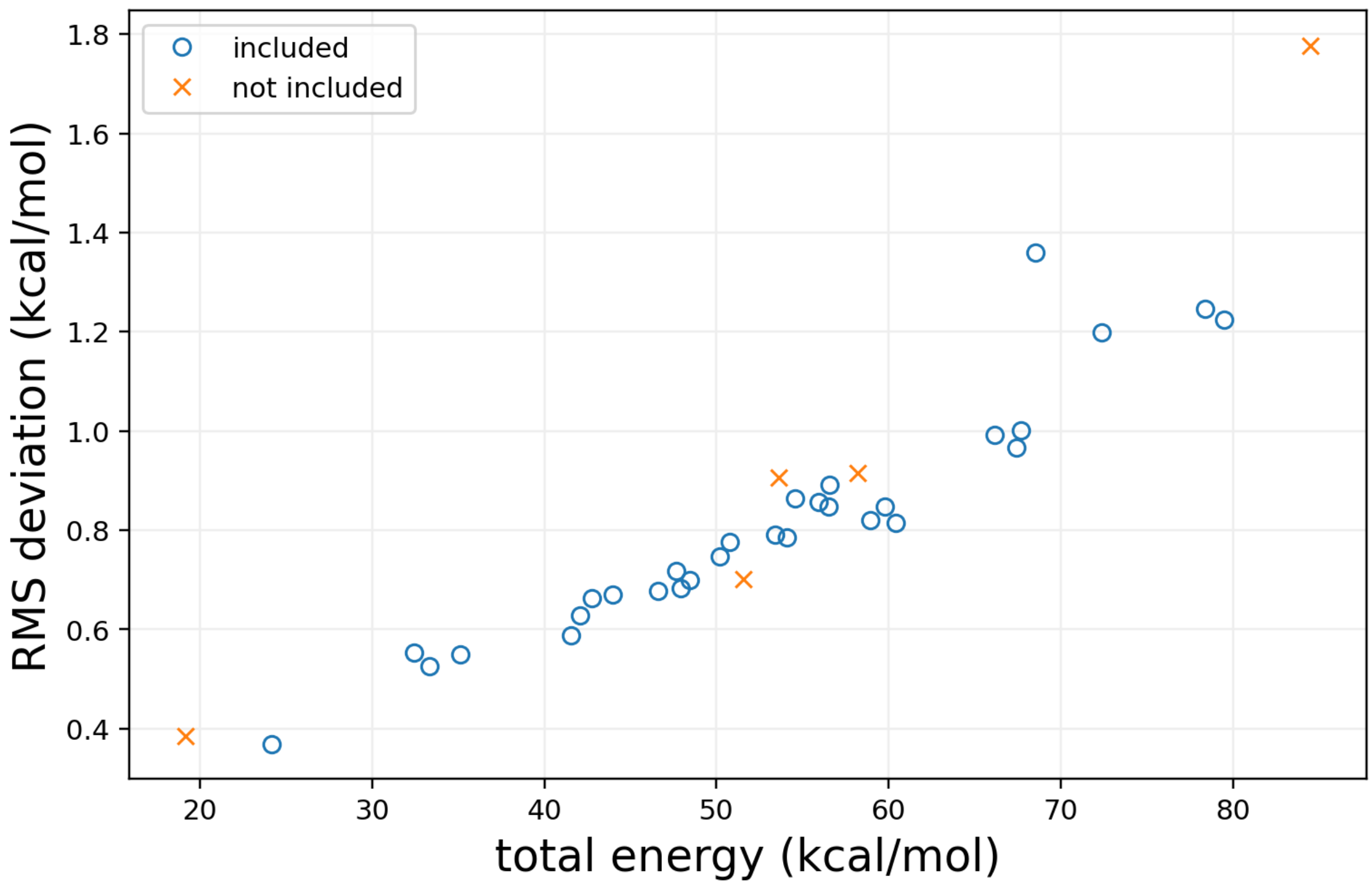

**Figure S7.** Root-mean-square deviation of the total energies of 10-ps trajectories propagated with the MDtrajNet model of ethanol. Total energy values present in the training set are marked with blue circles while values not included are marked with orange crosses. The energy values are relative to the global energy minimum.

***Tests on alanine***

For this, we first investigate whether the fine-tuning (transfer learning) of the foundational model can improve the data efficiency. As a test system, we take the alanine molecule comprising 13 atoms.

We first generated 32 NVE trajectories of 1-ps length using the ANI-1ccx potential energy surface and the conventional MD method. The initial structures were generated using Open Babeland initial velocities were sampled from a Maxwell–Boltzmann distribution at 300 K, with a time step of 0.05 fs.

Subsequently, we performed transfer learning from one of the foundational MDtrajNet-1 (m1–4) submodels, training the latter on 10,000, 50,000, 100,000, and 500,000 randomly sampled data points. Training metrics were recorded for each case. As a comparison, models were also trained from scratch using 100,000 and 500,000 data points.

We compare the validation loss evolution in transfer learning to that of the training MDtrajNet model from scratch (Figure S8).

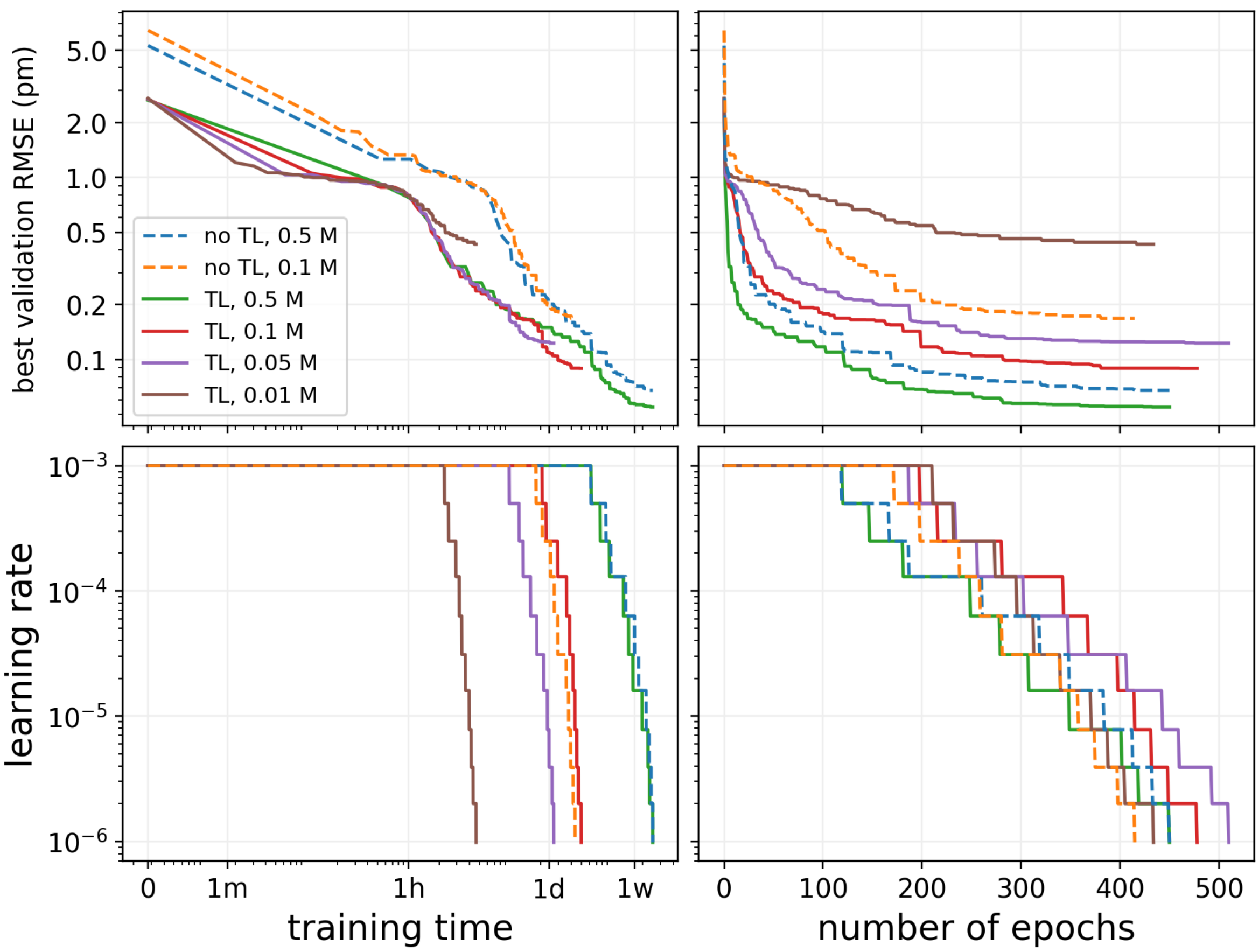

**Figure S8**. The training process of the alanine molecule. The best validation RMSE and learning rate versus both training time and number of epochs are shown. The models trained from scratch are shown

with dashed lines, and the one trained from a single MDtrajNet-1 submodel are shown with the solid lines. The sizes of training sets are shown in the legend. The log scale is used for the training time, with a manually labeled zero point for presenting the initial y-values. All models were trained with 4 CPU cores of Intel Xeon Silver 4210@2.2GHz and an Nvidia RTX 2080Ti GPU.

The results indicate that using the foundational model as a starting point leads to lower validation errors at the initial training stages compared to training from scratch. Throughout training, transfer learning consistently outperforms training from scratch, regardless of dataset size, when evaluated under the same training duration (upper left panel). In terms of final performance, the transferred models achieve lower errors at the same dataset size. Notably, transfer learning with just 50K points outperforms from-scratch training with 100K points. This suggests that the foundational model already possesses some predictive power for alanine.

This comparison demonstrates the transferability of the MDtrajNet architecture. Moreover, considering that the base model was trained on molecules with only 2–9 atoms, and alanine has 13 atoms, this result is very encouraging and indicates the possibility of obtaining universal models transferable to larger systems than those used for training.

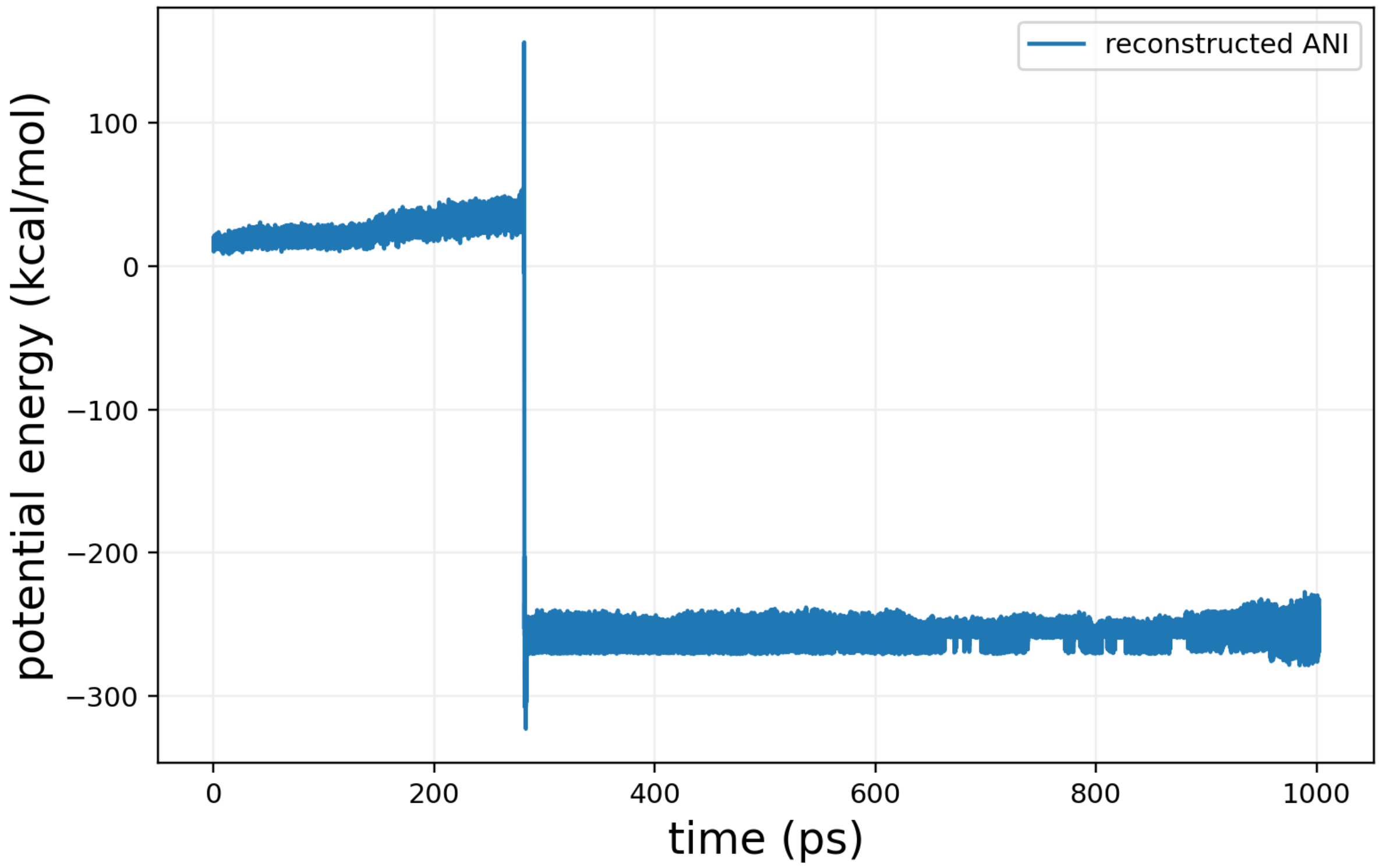

**Figure S9.** The potential energies in the 1-ns trajectory from the classical molecular dynamics using the reconstructed ANI model.

## *Additional figures and tables*

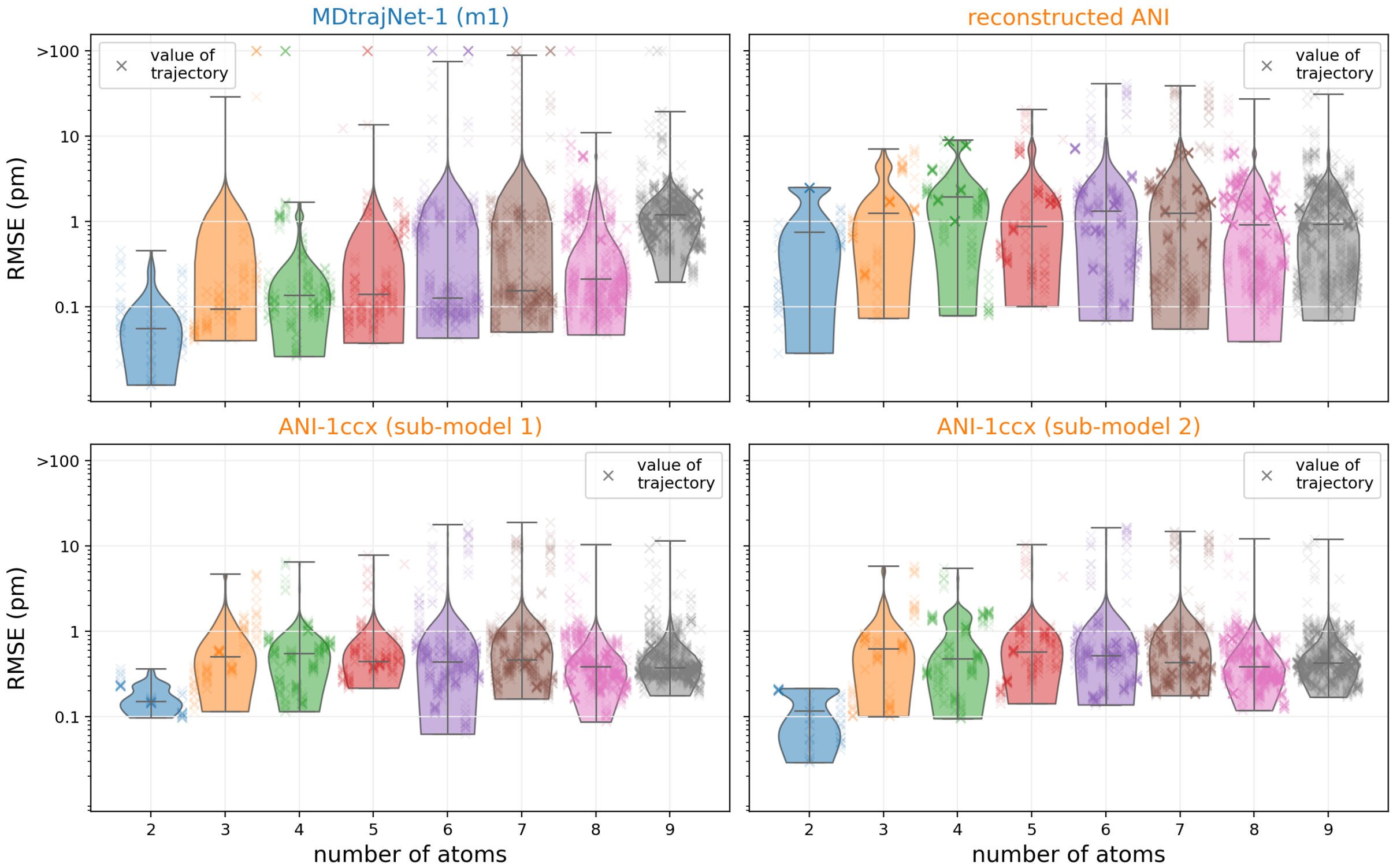

**Figure S10. Comparison of root-mean-square error of different models on the 200 test systems.** We take one sub-model from the foundational MDtrajNet-1 model to predict the nuclear coordinates as a function of time within the 10-fs time cutoff. The reconstructed ANI refers to the ANI model that was trained on the same data set as that used for training MDtrajNet-1. Results of each atomistic system are shown with an identical horizontal offset, which is further grouped by the number of atoms in the system. The violin plots show the distributions of RMSEs in groups (outliers are excluded), with extremum and median values marked. Results are shown for the test dataset.

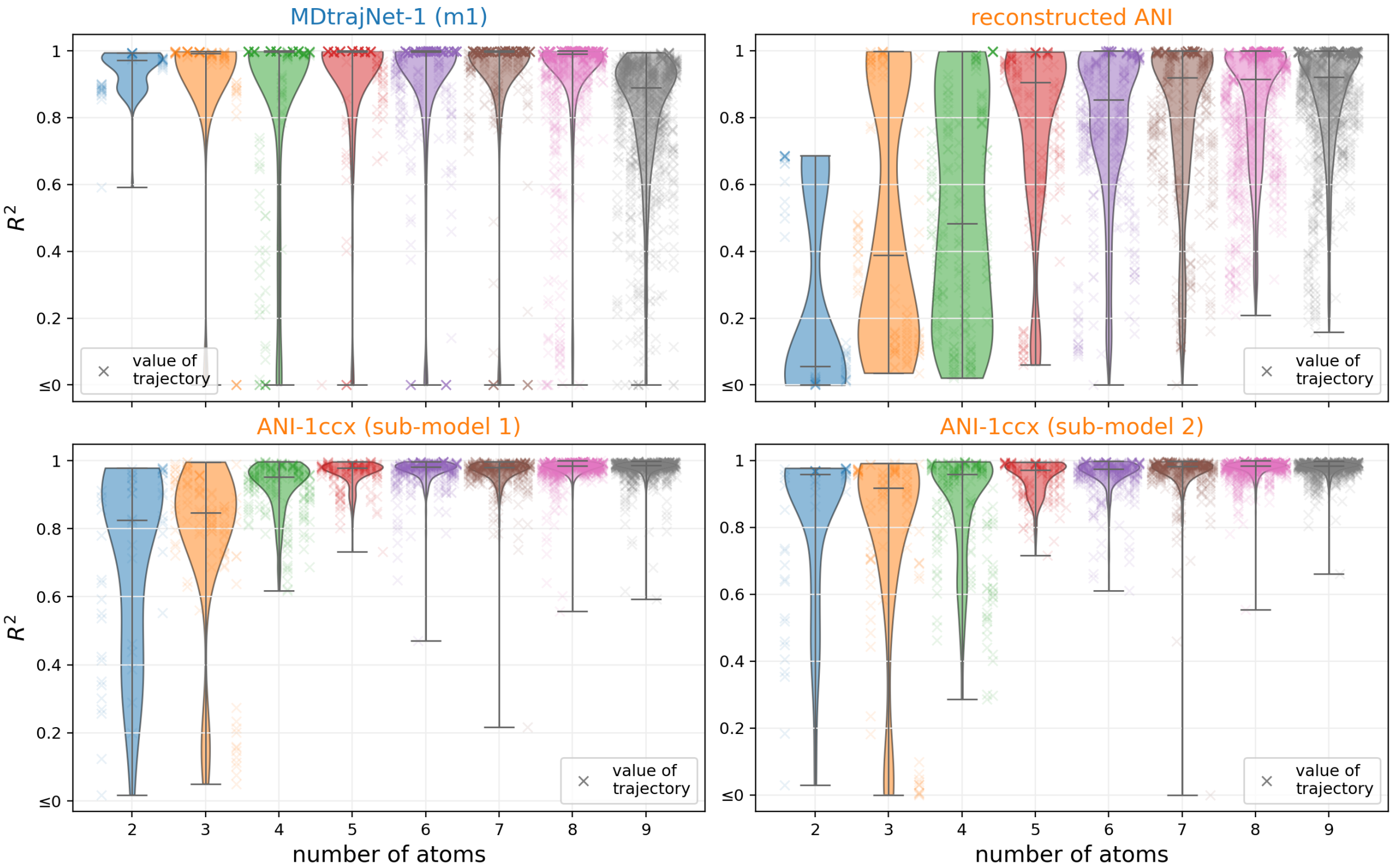

**Figure S11. Comparison of R-squares of different models on the 200 test systems.** We take one sub-model from the foundational MDtrajNet-1 model to predict the nuclear coordinates as a function of time within the 10-fs time cutoff. The reconstructed ANI refers to the ANI model that was trained on the same data set as that used for training MDtrajNet-1. Results of each atomistic system are shown with an identical horizontal offset, which is further grouped by the number of atoms in the system. The violin plots show the distributions of RMSEs in groups (outliers are excluded), with extremum and median values marked. Results are shown for the test dataset.

**Table S2.** Parameters of Lennard-Jones potential used in the periodic systems

| pair type | $\varepsilon$ (kcal/mol) | $\sigma$ (Å) |
|---|---|---|
| A-A | 0.02 | 0.5 |
| A-B | 0.04 | 0.6 |
| A-C | 0.09 | 1.2 |
| B-B | 0.08 | 0.9 |
| B-C | 0.12 | 1.6 |
| C-C | 0.20 | 2.0 |

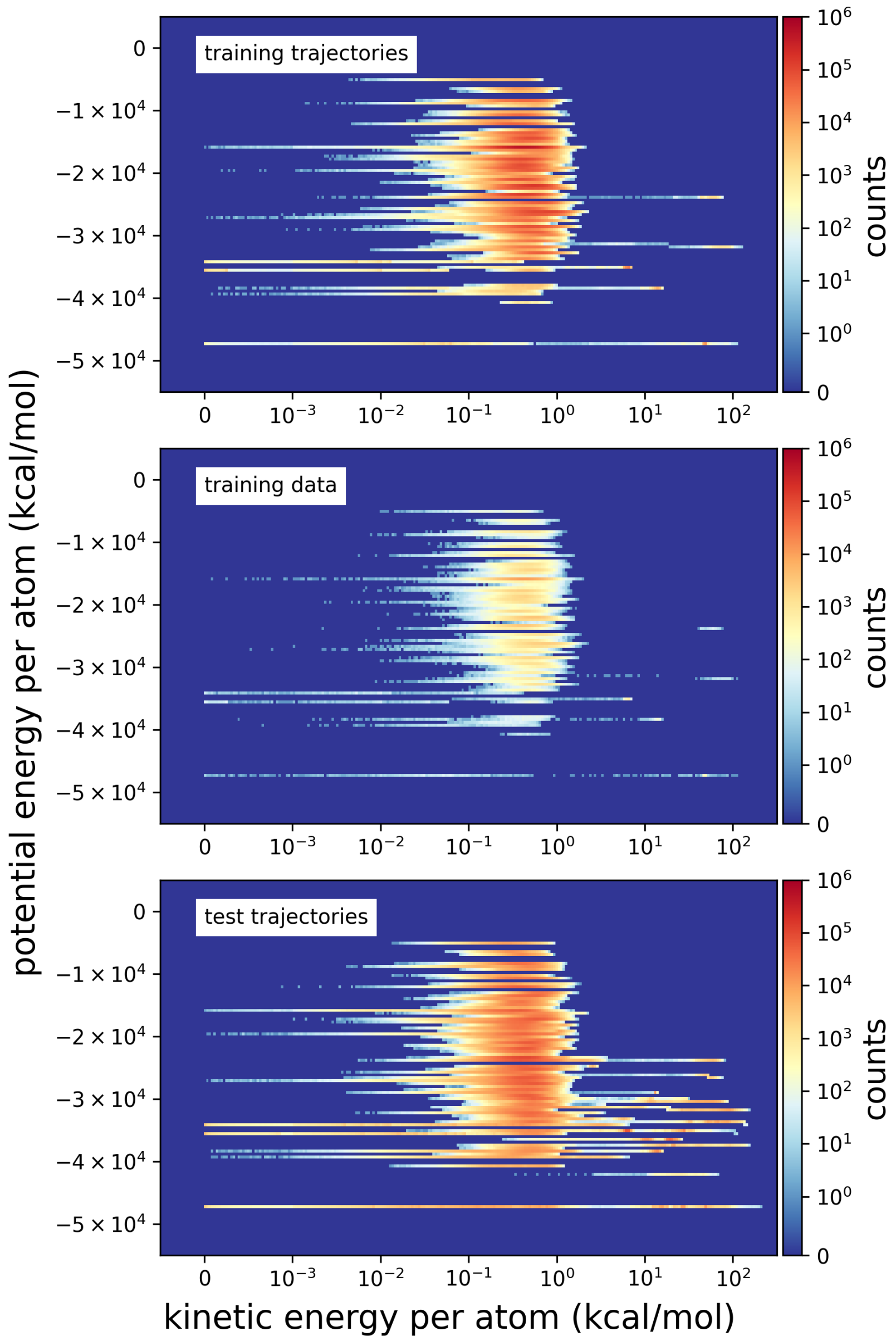

**Figure S12.** The distribution of data used in foundational MDtrajNet-1: top) All data in the training trajectories (43,784,378 points). middle) The data was randomly sampled from the training trajectories (1,000,000 points, ~2.28% of 43784378) as used for training. bottom) All data in the independent test trajectories (37,223,722 points), which has wider ranges in potential and kinetic energies.

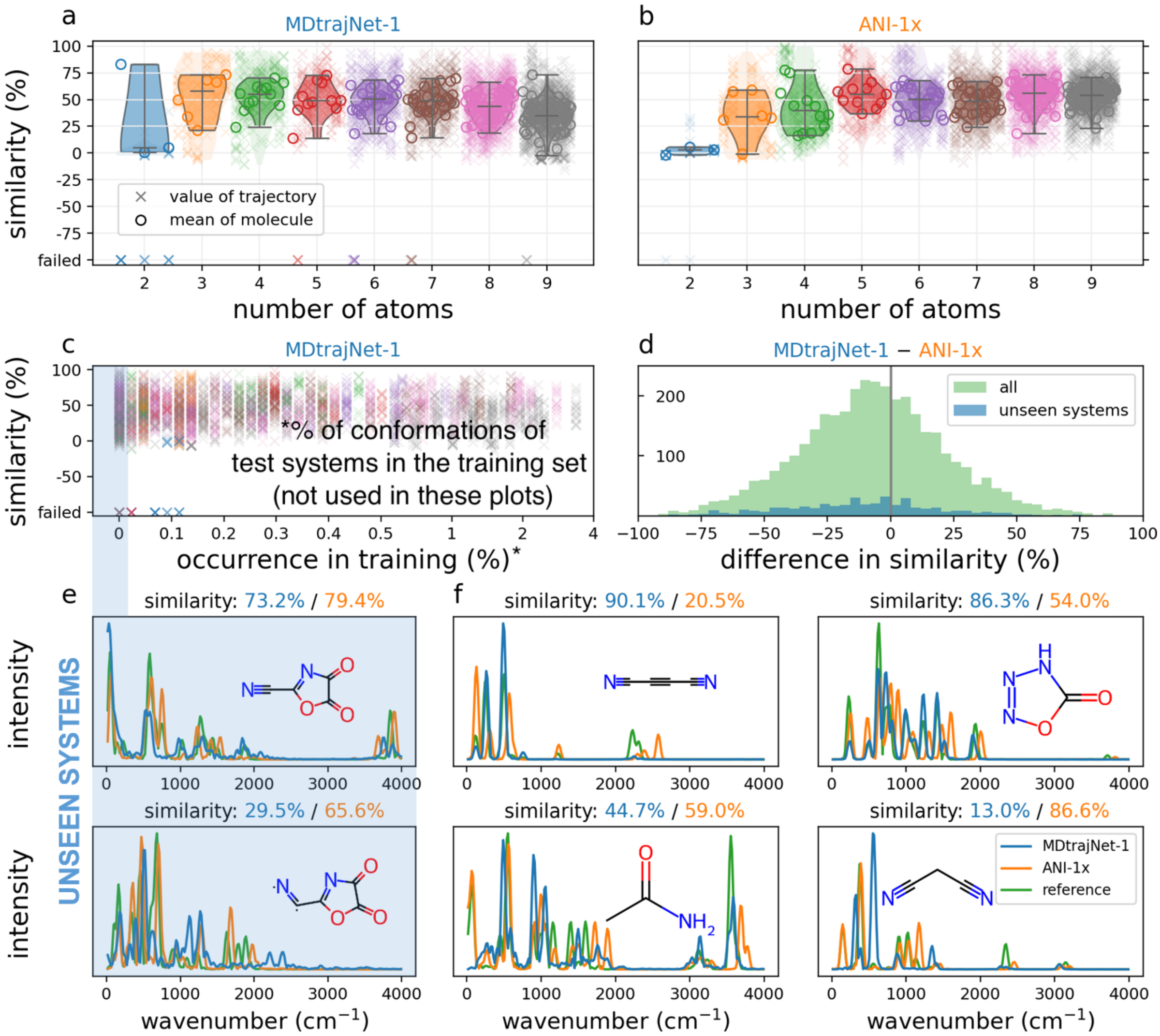

**Figure S13.** a, b) The similarity of power spectra for the 10-ps-long test trajectories generated with a) the foundational MDtrajNet-1 and b) the classical MD propagated with 0.05 fs time step using the universal ANI-1x potential of the DFT quality with respect to the reference. Each faint cross in the left plot represents the value for an individual test trajectory, while a circle represents the mean value of a tested molecule. Their distributions are also shown by the violin plots, whose ticks indicate the extreme and median values. Results of different systems are shown with identical horizontal offsets, which are further grouped by the number of atoms in the system. c) The spectral similarities of the test set versus the occurrence of the configurations of the corresponding test system in the training data (these training configurations are not used in any of the plots in this figure). d) The distribution of the differences in similarity in the stable trajectories of all and unseen molecules in the test data set. The difference is obtained by subtracting the similarity scores of ANI-1x from those of MDtrajNet-1 w.r.t. the reference ANI-1ccx trajectories. e, f) Representative spectra with varying similarities for unseen (e) and seen (f) molecules. Note that initial conditions for seen molecules were not included in the training set.

## References

(1) Ge, F.; Zhang, L.; Hou, Y.-F.; Chen, Y.; Ullah, A.; Dral, P. O., Four-Dimensional-Spacetime Atomistic Artificial Intelligence Models. *The Journal of Physical Chemistry Letters* **2023,** *14*, 7732-7743.